\documentclass[11pt,3p,number]{elsarticle}

\usepackage{amsmath,amsthm,amssymb,mathrsfs,enumerate,bbm,array}
\usepackage{graphicx,xcolor,url}
\usepackage{bm}

\renewcommand{\P}{\mathbb{P}}
\newcommand{\E}{\mathbb{E}}
\newcommand{\F}{\mathcal{F}}
\newcommand{\HH}{\mathcal{H}}
\newcommand{\OO}{\mathcal{O}}
\newcommand{\s}{\sigma}

\newcommand{\W}{\Omega}
\newcommand{\eps}{\epsilon}
\renewcommand{\l}{\lambda}
\newcommand{\R}{\mathbb{R}}

\DeclareMathOperator\logit{logit}

\DeclareMathOperator\sspan{span}

\newtheorem{lemma}{Lemma}

\theoremstyle{remark}
\newtheorem*{remark}{Remark}

\begin{document}
\begin{frontmatter}
\title{Statistical Emulators for Pricing and Hedging Longevity Risk Products}

\author{J. Risk and
	 M. Ludkovski \footnote{The authors are with the Department of Statistics \& Applied Probability, University of California, Santa Barbara CA 93106-3110;}
	 }

\begin{abstract}
We propose the use of statistical emulators for the purpose of valuing mortality-linked contracts in stochastic mortality models. Such models typically require (nested) evaluation of expected values of nonlinear functionals of multi-dimensional stochastic processes. Except in the simplest cases, no closed-form expressions are available, necessitating numerical approximation. Rather than building ad hoc analytic approximations, we advocate the use of modern statistical tools from machine learning to generate a flexible, non-parametric surrogate for the true mappings. This method allows performance guarantees regarding approximation accuracy and removes the need for nested simulation. We illustrate our approach with case studies involving (i) a Lee-Carter model with mortality shocks, (ii) index-based static hedging with longevity basis risk; (iii) a Cairns-Blake-Dowd stochastic survival probability model.
\end{abstract}

\begin{keyword}
%
Statistical emulation \sep longevity risk \sep life annuities \sep valuation of mortality-contingent claims
\end{keyword}

\end{frontmatter}

\section{Introduction}
Longevity risk has emerged as a key research topic in the past two decades. Since the seminal work of \citet*{lee1992modeling} there has been a particular interest in building stochastic models of mortality. Stochastic mortality allows for generation of a range of future longevity forecasts, and permits the modeler to pinpoint sources of randomness, so as to better quantify respective risk. Longevity modeling calls for a marriage between the statistical problem of calibration, i.e.~fitting to past mortality data, and the financial problem of pricing and hedging future longevity risk. At its core, the latter problem reduce to computing expected values of certain functionals of the underlying stochastic processes. For example, the survival probability for $t$ years for an individual currently aged $x$ can be expressed as a functional
\begin{equation}\label{eq:survival-prob}
P(t,x) = \E \left[ \exp \left( -\int_0^t \mu(s,x+s) \,ds \right) \right],
\end{equation}
where $\mu(s,x+s)$ is the force of mortality at date $s$ for an individual aged $x+s$. In the stochastic mortality paradigm $\mu(s,x+s)$ is random for $s > 0$, and so one is necessarily confronted with the need to evaluate the corresponding expectations on the right-hand-side of \eqref{eq:survival-prob}.

The past decade has witnessed a strong trend towards complexity in both components of \eqref{eq:survival-prob}. On the one hand, driven by the desire to provide faithful fits (and forecasts) to existing mortality data, increasingly complex mortality models for $\mu(t,x)$ have been proposed. The latest generation of models feature multi-dimensional, nonlinear stochastic state processes driving $\mu(\cdot,x)$, see e.g.~\citet{cairns2009quantitative,li2009uncertainty,lin2013pricing,Barrieu2012,fushimi2014bayesian}. These models are effective at calibration and emitting desirable forecasts, but lack tractability in terms of closed-form formulas. On the other hand, sophisticated insurance products, such as variable annuities or longevity swap derivatives, make valuation and hedging highly nontrivial, and typically call for  numerical approaches, as closed-form formulas are not available. Taken together, pricing of mortality-linked contracts becomes a complex system, feeding multi-dimensional stochastic inputs through a ``black box'' that eventually outputs net present value of the claim.


These developments have created a tension between the complexity of mortality models that do not admit explicit computations and the need to price, hedge and risk manage complicated contracts based on such models. Due to this challenge, there remains a gap between the academic mortality modeling and the implemented models by the longevity risk practitioners. Because the aforementioned valuation black box is analytically intractable, there is a growing reliance on Monte Carlo simulation tools, which in turn is accompanied by exploding computational needs. For example, many emerging problems require \emph{nested simulations} which can easily take days to complete. Similarly, many portfolios contain millions of heterogeneous products (see, e.g.~\citet{LinGan15}) that must be accurately priced and managed. In this article we propose to apply modern statistical methods to address this issue. Our approach is to bridge between the mortality modeling and the desired pricing/hedging needs through an intermediate \emph{statistical emulator}. The emulator provides a computationally efficient, high-fidelity surrogate to the actual mortality model. Moreover, the emulator converts a calibrated opaque mortality model into a user-friendly valuation ``app''. The resulting toolbox  allows a plug-and-play strategy, so that the end user who is in charge of pricing/risk-management can straightforwardly swap one mortality model for another, or one set of mortality parameters for an alternative. This modular approach allows a flexible solution to robustify the model-based longevity risk by facilitating comparisons of different longevity dynamics and different assumptions.

Use of \emph{emulators} is a natural solution to handle complex underlying stochastic simulators and has become commonplace in the simulation and machine learning communities \cite{Santner2003,kleijnen2007design}. Below we propose to apply such statistical learning within the novel context of insurance applications. In contrast to traditional (generalized) linear models, emulation calls for fully nonparametric models, which are less familiar to actuaries. To fix ideas, in this article we pursue the  problem of pricing/hedging vanilla life annuities, a foundational task in life insurance and pension plan management. Except in the simplest settings, there are no explicit formulas for annuity values and consequently approximation techniques are already commonplace. Looking more broadly, our method would also be applicable in many other actuarial and risk management contexts, see Section~\ref{sec:conclusion}.

The paper is organized as follows: In Section \ref{sec:MainObjective} we introduce the emulation problem and review the mathematical framework of stochastic mortality. Section \ref{sec:emulation} discusses the construction of emulators, including spline and kriging surrogates, as well as generation of training designs and simulation budgeting. The second half of the paper then presents three extended case studies on several stochastic mortality models that have been put forth in the literature. In Section \ref{sec:chen-coxcasestudy} we examine a Lee-Carter model with mortality shocks that was proposed by Chen and Cox \cite{chen2009modeling}; Section \ref{sec:2pop-study} studies approximation of hedge portfolio values in a two-population model based on the recent work by Cairns et al \cite{cairns2014longevity}. Lastly, Section \ref{sec:cooked} considers valuation of deferred annuities under a Cairns-Blake-Dowd (CBD) \cite{cairns2006two} mortality framework.

\section{Emulation Objective}\label{sec:MainObjective}

We consider a stochastic system with Markov state process $Z=(Z(t))$.  Throughout the paper we will identify $Z$ with the underlying stochastic \emph{mortality factors}. In Section \ref{sec:stoch-mortality} we review some of the existing such models and explicit the respective structure of $Z$. Typically, $Z$ is a multivariate stochastic process based on either a stochastic differential equation or time-series $ARIMA$ frameworks. For example, $Z$ may be of diffusion-type or an auto-regressive process.

In the inference step, the dynamics of $Z$ are calibrated to past mortality data that reflect as closely as possible the population of interest. In the ensuing valuation step, the modeler seeks to evaluate certain quantities related to a functional $F(T,Z(\cdot))$ looking into the future. The time horizon $T \ge 0$ allows consideration of deferred contracts that are common in longevity risk, see below.
Our notation furthermore indicates that $F$  potentially depends on the whole path $\{Z(t), t \ge T\}$, such as \begin{equation}
F(T,Z(\cdot)) = \exp \Bigl( -\sum_{t=T}^\infty h(Z(t)) \Bigr), \end{equation}
 for some $h(z).$  Given $F$, the most common aim is to compute its expected value, based on the initial data at $t=0$,
\begin{align}
\label{eq:expected-F}
\E \left[ F(T,Z(\cdot)) \mid Z(0) \right].
\end{align}
Other summary statistics of interest in actuarial applications include the
\begin{itemize}

    \item Quantile $q(\alpha; F(T,Z(\cdot)))$ (eg.~the Value-at-Risk at level $\alpha$ of $F$);

    \item Expected Shortfall of $F$, $\E[F(T,Z(\cdot)) \mid F(T,Z(\cdot)) \leq q(\alpha; F(T,Z(\cdot))), Z(0)]$;

  \item Correlation between two functionals, $Corr( F_1(T, Z(\cdot)), F_2(T,Z(\cdot)) | Z(0))$.

\end{itemize}

To fix ideas we henceforth focus on \eqref{eq:expected-F} which is a fundamental quantity in pricing/hedging problems. When $T > 0$, the evaluation of \eqref{eq:expected-F} can be broken into two steps, namely first we evaluate
\begin{align}\label{def:f}
f(z) \doteq \E[ F(T,Z(\cdot)) | Z(T) = z],
\end{align}
and then use the Markov property of $Z$ to carry out an outer average,
$$\E[ F(T,Z(\cdot)) | Z(0)] = \int_{\mathbb{R}^d} f(z) p_{T}(z| Z(0)) dz,$$ where $p_{T}(z'|z) = P( Z(T)=z' | Z(0)=z)$ is the transition density of $Z$ over $[0,T]$.

Crucially, because the form of $F(T,Z(\cdot))$ is nontrivial, we shall assume that $f(z)$ is not available explicitly, and there is no simple  way to describe its functional form. However, since $f(z)$ is a conditional expectation, it can be sampled using a simulator, i.e.~the modeler has access to an engine that can generate independent, identically distributed samples $F(T,Z^{(n)}(\cdot))$, $n=1,\ldots,$ given $Z(0).$ However this simulator is assumed to be expensive, implying that computational efficiency is desired in using it.

 Given an initial state $Z(0)$, a naive Monte Carlo approach to evaluate \eqref{eq:expected-F} is based on nested simulation. First, the outer integral over $p_T(z| Z(0))$ is replaced by an empirical average of \eqref{def:f} across $m=1,\ldots,N_{out}$ draws $z^{(m)} \sim Z(T) | Z(0)$,
\begin{align}\label{eq:N-out}
  \E[ F(T,Z(\cdot)) | Z(0)] \simeq \frac{1}{N_{out}} \sum_{m=1}^{N_{out}} f(z^{(m)}).
\end{align}
Second, for each $z^{(m)}$ the corresponding inner expected value $f(z^{(m)})$ is further approximated via
\begin{equation}\label{eq:N-in}
f(z^{(m)}) \simeq \frac{1}{N_{in}} \sum_{n=1}^{N_{in}} F(T,z^{(m),n}(\cdot)), \qquad m = 1, \ldots, N_{out},
\end{equation}
where $z^{(m),n}(t), t \geq T$ are $N_{in}$ independent trajectories of $Z$ with a fixed starting point $z^{(m),n}(T) = z^{(m)}$. This nested approach offers an unbiased but expensive estimate. Indeed, the total simulation budget is $\OO(N_{out} \cdot N_{in})$ (where the usual big-Oh notation $h(x)=\OO(x)$ means that $h(\cdot)$ is asymptotically linear in $x$ as $x\to\infty$)  which can be computationally intensive  -- for example 1,000 simulations at both steps requires $10^6$ total simulations.

For this reason, it is desirable to construct cheaper versions of approximating \eqref{eq:expected-F}. The main idea is to replace the inner step of repeatedly evaluating $f(z)$ (possibly for some very similar values of $z$) with a simpler alternative. One strategy is to construct deterministic approximations to \eqref{def:f} by replacing the random variable $Z(s)|Z(T)$, $s > T$ with a fixed constant, e.g.~its mean, which can then be plugged into $F$ to estimate the latter's expected value. This effectively removes the stochastic aspect and allows to obtain explicit approximations to $f(\cdot)$. (The simplest approximation is to simply freeze $Z(s) = Z(T) \forall s > T$.) However, the resulting error is hard to judge, and moreover, analytic, off-line derivations are needed to obtain a good approximation. Consequently, we advocate the more statistical method of utilizing a \emph{surrogate} model for $f(\cdot)$. This approach can be generically used in any Markovian setting, requires no analytic derivations, and makes minimal a priori assumptions about the structure of $f(\cdot)$.

The main idea of emulation is a regression framework that generates a fitted $\hat{f}(\cdot)$ by solving regression equations over a training dataset $\{ z^{(n)}, F(T, z^{(n)}(\cdot)) \}_{n=1}^{N_{tr}}$ of size $N_{tr}$. Emulation reduces approximating $f(\cdot)$ to the twin statistical problems of (i) experimental design (generating the training dataset) and (ii) regression (specifying the optimization problem that the approximation $\hat{f}$ solves). Details of these steps are presented in Section \ref{sec:emulation} below.

Because we are fitting a full response model, rather than a pointwise estimate, the emulator budget $N_{tr} \gg N_{in}$ will be an order of magnitude bigger than in \eqref{eq:N-in}. It will also require regression overhead. However, once $\hat{f}$ is fitted, prediction of $\hat{f}(z)$ for a particular value $z$ takes $\OO(1)$ effort, so that we can use \eqref{eq:N-out} to estimate the original problem in \eqref{eq:expected-F} at a cost linear in $N_{out}$. To sum up, the total budget of the emulator is just $\OO(N_{tr} + N_{out})$, much smaller than $\mathcal{O}( N_{out} \times N_{in})$ of nested Monte Carlo. These savings become even more significant as the dimension of state $Z$ grows. Indeed, with multi-dimensional models, both $N_{out}$ and $N_{in}$ need to be larger to better cover the respective integrals over $\R^d$, and hence the efficiency of nested simulations will deteriorate quickly. Intuitively, the latter computational budget is at least quadratic in $d$. In contrast, the intuitive complexity of an emulator is linear in $d$. As stochastic mortality models become more complex,  models with $d=3,4,5+$ factors are frequently proposed, and efficiency issues become central to the ability of evaluating \eqref{eq:expected-F} tractably.




\subsection{Valuation of Life Annuities}
In longevity modeling, $Z$ represents the stochastic factors driving the central force of mortality $m(t,x)$. Formally, $Z = (Z(t)) = (Z_1(t), \ldots, Z_d(t))$ is a $d-$dimensional $(\F(t))$ measurable Markov process on a complete filtered probability space $(\W, \F, \P, (\F(t)))$. The filtration  $(\F(t))$ is the information up to time $t$ of the evolution of the mortality processes.

A typical state-of-the-art model decomposes $m(t,x)$ into a longevity trend, an Age effect, and a Cohort effect (known collectively as APC models). Each of the above may be modeled in turn by one or more stochastic factors.  The most common models are the Lee-Carter \cite{lee1992modeling} and CBD \cite{cairns2006two} models and their generalizations.  Generally their individual components follow an ARIMA model; details can be found in the survey \citet{cairns2009quantitative}.  

To deal with cashflows at different dates, we assume the existence of a risk-free asset and denote by $B(T,T+s)$ the price of an $s$-bond at date $T$ with maturity at $T+s$. For the rest of the article we will assume constant force of interest $r$, leading to $B(T,T+s) = e^{-r s}$. One can straightforwardly handle stochastic interest rates (which then form part of $Z(\cdot)$); see \citet{jalen2009valuation} for a discussion of correlation structure between mortality and interest rates and \citet{fushimi2014bayesian} for an example that applies Bayesian methods to longevity derivative pricing under a Cox-Ingersoll-Ross interest rate model. 

Consider an individual aged $x$ at time 0 whose remaining lifetime random variable is denoted as $\tau_x.$  The state process $Z$ captures $m(t,x+t),$ the mortality rate  process for $\tau_x$ at time $t$, when the individual would have aged to $x+t$.
  For small $dt,$ the instantaneous probability of death is approximately $m(t,x+t)dt,$ so that  
the random survival function of $\tau_x$ is
\begin{equation}
S(t,x) \doteq \exp\left(-\int_0^t m(s,x+s) \right).
\end{equation}
More generally for $u \le t < T$,  the probability of an individual aged $x$ to survive between dates $t$ and $T$, given the information at time $u$ is given by
\begin{align} \label{eq:P}
\P(\tau_x > T \mid \tau_x > t, \F_u) 
	&= \E\left[\left. \frac{S(T,x)}{S(t,x)} \right| \F_u \right]\\ \notag
	&= \E\left[ \left. \exp\left(-\int_t^T m(s,x+s)\right) \right| Z(u) \right] \doteq P(Z(u); t, T, x),
\end{align}
where the last equality follows from the Markov property.  
The deterministic analogue of $P(Z(0); t,T,x)$ in actuarial literature is ${}_{T-t}p_{x+t}.$

As a canonical actuarial contract, we henceforth focus on deferred life annuities. These contracts are fundamental to valuation of defined benefit pension plans, which normally begin paying annuitants at retirement age (typically age 65) and continue until their death, possibly with survivor benefits. (For valuation purposes the payment is assumed to end  at some pre-specified upper age $\bar{x}$, e.g.~100 or 110).   A major problem of interest is valuing such life annuities for current plan participants who are still working, i.e.~under age 65. Because this requires making longevity projections many decades into the future, longevity risk becomes a crucial part of risk management. The net present value of a life annuity at date $T$ is
\begin{align}\label{eq:life-annuity}
  a(Z(T);T,x) \doteq \sum_{s=1}^\infty B(T,T+s) \P(\tau_x \geq T+s \mid \F_T) = \sum_{s=1}^{\bar{x}-x} e^{-r s} P(Z(T); T, T+s, x),
\end{align}
where we emphasize that the random mortality shocks come from $Z$. Finally, the net present value at $t=0$ is
$NPV \doteq \E[ e^{-r T} \cdot a(Z(T),T,x)]$, which can be seen as an instance of \eqref{eq:expected-F} that includes discounting and integrating over the density of $Z(T)$.
Except for the simplest models, the survival probability $P(z; \cdot)$ is not analytically known and hence neither is \eqref{eq:life-annuity} or NPV. Without a representation for $z \mapsto a(z,T,x)$ one is then forced to resort to approximations for all the basic tasks of pricing, hedging, asset liability management, or solvency capital computation. The discussed nested simulation takes the form of first approximating $a(z^{(1)},T,x)$ for some representative scenarios $(z^{(1)}, \ldots, z^{(n)})$, and then further manipulating the resulting ``empirical'' distribution of $(a(z^{(1)},T,x), \ldots, a(z^{(n)},T,x))$.  Emulation provides a principled statistical framework for optimizing, assessing and improving such two-level simulations.

\begin{remark}
  As mentioned, estimation of $a(\cdot, T,x)$ is usually a building block embedded in a larger setting which requires repeated evaluation of the former quantity. For instance, \citet{bauer2012calculation} addresses nested Monte Carlo simulations in calculating the present value of life-annuity-like instruments in the calculation of solvency capital requirements. Let us also mention the works \citet{BacinelloBiffis10,BoyerStentoft13} who considered valuation of mortality contracts with early exercise features, such as surrender guarantees. The respective least squares Monte Carlo algorithms can be seen as classical parametric linear-model emulators in our terminology, eschewing the need for nested Monte Carlo forests.
\end{remark}

%
%

\subsection{Stochastic Mortality}\label{sec:stoch-mortality}

We concentrate on discrete-time mortality models which are easier to calibrate to the discrete mortality data, typically aggregated into annual intervals. The common assumption is that the central force of mortality remains constant through a given calendar year, so that for all $0 \leq s, u \leq 1,$ we have $m(t+s, x+u) = m(t,x)$.
%
Therefore
\begin{align}
P(Z(u);t,T,x) &= \E\left[ \left. \exp\left(-\sum_{s=t+1}^T m(s,x+s)\right) \right| Z(u) \right], \quad u \le t < T.\label{eq:P2}
\end{align}
Thus, $P(Z(T); T, T+s, x+T)$ becomes a functional of the trajectory of $Z$.

Three major approaches to stochastic mortality have been put forward in the literature. The first approach, pioneered by \citet{lee1992modeling}, directly treats $m(t,x)$ as a product of individual stochastic processes, e.g.~$ARIMA$ time-series.  This setup allows incorporating demographic insights, as well as disentangling age, period and cohort effects in future forecasts. To wit, the popular age-period-cohort (APC) mortality models assume that (see \ref{sec:leecarter-cbd} for more details)
\begin{align}
  \label{eq:m2} \log m(t,x) = \beta_x^{(1)} + \frac{1}{n_a} \kappa^{(2)}(t) + \frac{1}{n_a}  \gamma^{(3)}(t-x),
\end{align}
where $\kappa^{(2)}$ and $\gamma^{(3)}$ are stochastic processes and $n_a$ is the number of ages that $x$ can take in fitting.  In this case, the state process $Z(t)$ depends on current and potentially past values of $\kappa^{(2)}$ and $\gamma^{(3)}.$
Attempts to understand the statistical validity of such models have been done by, for example, \citet{lee2001evaluating}, \cite{brouhns2002poisson}, \citet{booth2002applying}, \citet{czado2005bayesian}, \citet{delwarde2007smoothing}, and \citet{li2009uncertainty}.  In addition, there have been several extensions of the Lee Carter model by \citet{renshaw2006cohort}, \citet{hyndman2007robust}, \citet{plat2009stochastic}, \citet{debonneuil2010simple}, and \citet{cairns2011bayesian}.

None of these models admit closed form expressions for survival probabilities $P(z; \cdot)$. Consequently, several authors have proposed approximation methods. \citet{coughlan2011longevity} used a bootstrapping approach, while \citet{cairns2014longevity} derived an analytic approximation, commenting that industry practice is to utilize deterministic projections. The more flexible tool of Monte Carlo simulation has been applied in \citet{bauer2012calculation} among others.


The second approach, due to \citet{cairns2006two} (CBD), generates a stochastic model for the survival probability \eqref{eq:P2}, allowing for straightforward pricing of longevity-linked products; however, it is more difficult to calibrate and to obtain reasonable forecasts for future mortality experience in a population as a whole. The third approach works with forward mortality rates \cite{bauer2012modeling}, borrowing ideas from fixed income markets. Forward models give a holistic view of how the mortality curves can evolve over time, and presents a dynamically consistent structure for mortality forecasting. Once again however, they do not provide easy expressions for \eqref{eq:P2} and hence require further manipulation for pricing purposes.


\subsection{Bias/Variance Trade-Off}
With a view towards approximating \eqref{eq:life-annuity}, it is imperative to first quantify the resulting quality of an approximation. The standard statistical approach is to use the framework of mean squared error.
Fix $z$ and let $a(z) \equiv a(z, T,x)$ be the true value of a life annuity conditional on state $Z(T)=z$. If $a(z)$ is  being estimated by $\hat{a}(z),$ then
\begin{equation}\label{eq:IMSEBias}
\text{IMSE}(\hat{a}) \doteq \E\left[(\hat{a}-a)^2\right],  \qquad \text{Bias}(\hat{a}) = \E\left[\hat{a}-a\right],
\end{equation}
where the averaging is over the sampling distribution (i.e.~different realizations of data used in constructing it) of $\hat{a}(z)$.

Starting with \eqref{eq:IMSEBias} leads to the fundamental bias/variance trade-off. At one end of the spectrum, a Monte Carlo estimate as in \eqref{eq:N-in} has zero bias but carries a high variance. At the opposite end, an analytic approximation has zero variance, but will have a non-zero bias that cannot be alleviated (whereas the Monte Carlo IMSE will go to zero as the size of the dataset grows $N_{tr} \to \infty$) even asymptotically. Because low variance is often preferred practically, analytic methods have remained popular. \citet{cairns2014longevity} echoes that it is usual practice in industry to use a deterministic projection of mortality rates rather than use a simulation approach.  

 The basic idea for the deterministic approximations is that if $\hat{m}(t,x)$ is an unbiased estimate for $m(t,x),$ then
\begin{align}
P(Z(u); t, T, x) &=  \E\left[ \left. \exp\left(-\sum_{s=t+1}^T m(s,x+s)\right) \right| Z(u) \right] \notag\\
	&\approx \exp\left(-\sum_{s=t+1}^T \E\left[m(s,x+s)\mid Z(u)\right]\right)
	= \exp\left(-\sum_{s=t+1}^T \hat{m}(s,x+s; Z(u))\right). \label{eq:LCapprox}
\end{align}
Using the estimate for $P(Z(u); \cdot)$ in \eqref{eq:LCapprox} one can then approximate $a(Z(T), T,x )$ term-wise.
Jensen's inequality implies that $\exp\left(-\sum_{s=t+1}^T \hat{m}(s,x+s)\right) > P(Z(u); t, T, x)$. Consequently, any such approximation is guaranteed to be biased high for the survival probabilities (and subsequently the annuity values).

Analytic approximations can be very powerful and of course very fast but they carry two major disadvantages. One is the need to \emph{derive} a suitable estimator $\hat{m}$. This may be possible in a simple model (e.g.~low-dimensional $Z$ with linear dynamics, like in the original Lee-Carter model), but otherwise may require a lot of off-line labor, leading to unnecessary focus on simplifications at the expense of calibration and risk management consistency. Second, the  degree of accuracy of the approximation is unknown. Indeed, there is generally not much that is available about empirical accuracy of the right-hand-side in \eqref{eq:LCapprox} for a given model, leaving the user in the dark about how much error is being made. This issue is very dangerous, since potentially major mis-valuations may creep up unbeknownst to the risk manager.

To remedy the above shortcomings, while still maintaining significant variance reduction compared to plain MC, we advocate the use of statistical emulators. The latter offer posterior quantification of accuracy (via standard error or Bayesian posterior variance), and do not require any simplifications of the mortality model. An additional advantage is that one can directly approximate $z \mapsto a(z,T,x)$ without having to do intermediate approximations of the survival probabilities (which inevitably lead to further error compounding).  As we demonstrate in the case studies, statistical models for $a(z)$ can indeed efficiently address the bias/variance trade-off, by maintaining negligible bias and small variance, leading to improved IMSE metrics compared to other approaches.

%
%

\section{Statistical Emulation}\label{sec:emulation}
The idea of emulation is to replace the computationally expensive process of running a Monte Carlo sub-routine to evaluate $f(z)$ for each new site $z$ with a cheap-to-evaluate surrogate model that statistically predicts $f(z)$ for any $z \in \mathbb{R}^d$ based on results from a training dataset. At the heart of emulation is statistical learning. Namely, the above predictions are based on first obtaining pathwise estimates $y^{(n)} = F(T, z^{(n)} )$, $n=1,\ldots,N_{tr}$ for a set of training locations, called a design $\mathcal{D} \doteq (z^{(1)}, \ldots, z^{(N_{tr})})$. Next, one regresses $\{y^{(n)}\}$ against $\{z^{(n)}\}$ to ``learn'' the response surface $\hat{f}(\cdot)$. The regression aspect allows to borrow information across different scenarios starting at various sites. This reduces computational budget compared to the nested simulation step of independently making $N_{tr}$ pointwise estimates  $f(z^{(n)})$ by running $N_{in}$ scenarios from \emph{each} site $z^{(n)}$. The conceptual need for regression is two-fold. First, the emulator is used for interpolation, i.e.~using existing design to make predictions at new  sites $z$. In contrast, plain Monte Carlo only predicts at $z^{(n)}$'s. Second, like in the classical approach, the emulator \emph{smoothes} the Monte Carlo noise from sampling trajectories of $\{Z(s), s > T\}$.


Formally, the statistical problem of emulation deals with a sampler (or oracle)
\begin{align}\label{eq:oracle}
  Y(z) = f(z) + \eps(z),
\end{align}
where we identify $f(z) \equiv a(z, T,x)$ with the unknown \emph{response surface} and $\eps$ is the sampling noise, assumed to be independent and identically distributed across different calls to the oracle.  We make the assumption $\eps(z) \sim N(0, \tau^2(z)),$ where $\tau^2(z)$ is the sampling variance that depends on the location $z$. Emulation now involves the (i) experimental design step of proposing a design $\mathcal{D}$ that forms the training dataset, and (ii) a learning procedure that uses the queried results $(z^{(n)},y^{(n)})_{n=1}^{N_{tr}}$, with the $y^{(n)}$ being realizations of \eqref{eq:oracle} given $z^{(n)}$, to construct a fitted response surface $\hat{f}(\cdot)$. The fitting is done by specifying the approximation function class $\hat{f} \in \HH$, and a loss function $L(\hat{f}, f)$ which is to be minimized. The loss function measures the relative accuracy of $\hat{f}$ vis-a-vis the ground truth; in this paper we focus on the mean-squared approximation error
\begin{align}\label{eq:MSAE}
  L(\hat{f}, f) \doteq \int_{\mathbb{R}^d}  |\hat{f}(z) - f(z)|^2 dz.
\end{align}
Because the true $f$ is unknown, the definition of $L(\hat{f}, f)$ cannot be operationalized and instead a proxy based on the uncertainty (such as Bayesian posterior uncertainty or standard errors) surrounding $\hat{f}$ is applied. Also, since the structure of $f$ is unknown, it is desirable that the approximation class $\HH$ is dense, i.e.~has a sufficiently rich architecture to approximate any $f$ to an arbitrary degree of accuracy. To this end, we concentrate on kernel regression methods, namely linear smoothers. In the next subsections we introduce two such regression families, smoothing splines and kriging (Gaussian process) models.

\begin{remark}
  In this paper we focus on the original task of producing an accurate approximation to $f$ everywhere. In some contexts, accuracy is judged not globally, but locally, so that a differentiated accuracy measure is used. For example, in VaR applications, the model for $f$ must be accurate in the left-tail, but can be rather rough in the right-tail. In this case, \eqref{eq:MSAE} can be replaced by a weighted loss metric.
\end{remark}


\subsection{Emulators based on Spline Models}\label{sec:splines}
We generate emulators $\hat{f}(\cdot)$ using a regularized regression criterion. To wit, given a smoothing parameter $\l \ge 0$ we look for the minimizer $\hat{f} \in \HH$ of the following penalized residual sum of squares problem
\begin{align}\label{eq:spline}
RSS(f, \l) = \sum_{n=1}^{N_{tr}} \{y^{(n)} - f(z^{(n)})\}^2 + \l J(f),
\end{align}
where $J(f)$ is a penalty or regularization function. We concentrate on the case where the approximation class has a reproducing kernel Hilbert space (RKHS) structure which also generates $J(f)$. Namely, there exists an underlying positive definite kernel $C(z,z')$  such that $\HH_C = \sspan ( C(\cdot, z) : z\in \mathbb{R}^d )$ is the Hilbert space generated by $C$ and $J(f) = \| f \|^2_{\HH_C}$. The representer theorem implies that the minimizer of \eqref{eq:spline} has an expansion in terms of the eigen-functions
\begin{equation}\label{eq:reg-regression}
\hat{f}(z) = \sum_{j=1}^{N_{tr}} \alpha_j C(z,{z}^{(j)}),
\end{equation}
relating the prediction at $z$ to the kernel function sampled at the design sites $z^{(j)}$.

Our first family are smoothing (or thin-plate) splines that take \begin{equation}\label{eq:2d-tpsconstraint}
 J(f) = \int_{\R^d} \left[ \sum_{i,j=1}^d \frac{\partial}{\partial z_i}\frac{ \partial}{\partial z_j} f( z)\right] d z,
\end{equation}
and $\HH$ as the set of all twice continuously-differentiable functions. It is known \cite[Chapter 5]{hastie2009elements} that in this case the underlying kernel is given by $C(z, z') = \| z - z'\|^2 \log \| z - z' \|, $
where $\|\cdot\|$ denotes the Euclidean norm in  $\R^d$. The resulting optimization of \eqref{eq:spline} along with \eqref{eq:2d-tpsconstraint} gives a smooth response surface which is called a thin-plate spline (TPS), and has the explicit form
\begin{equation}\label{eq:tps-solution}
f(z) = \beta_0 + \bm{\beta}^T \vec{z} + \sum_{j=1}^{N_{tr}} \alpha_j \|z-z^{(j)}\|^2 \log \|z-z^{(j)}\|,
\end{equation}
with $\bm{\beta} = (\beta_1, \dots, \beta_d)^T$.

In 1-d, the penalized optimization reduces to \begin{align}\label{eq:1d-spline}
\inf_{f \in \mathcal{C}^2} \sum_{i=1}^{N_{tr}} \{y^{(n)} - f(z^{(n)})\}^2 + \l \int_\R \{f''(u)\}^2 du.
\end{align}
The summation in \eqref{eq:1d-spline} is a measure of closeness of data, while the integral penalizes the fluctuations of $f$.  Note that $\l=\infty$ reduces to the traditional least squares linear fit $\hat{f}({z}) = \beta_0 + \beta_1 {z}$ since it introduces the constraint $f''({z}) = 0.$ It is well known that the resulting solution is an expansion in terms of natural cubic splines,
i.e.~$\hat{f}$ is a piecewise cubic polynomial that has continuous first and second derivatives at the design sites $z^{(n)}$, and is linear outside of the design boundary.
%

Several methods are available to choose the smoothing parameter $\l$, including cross-validation or MLE \citet[Chapter 5]{hastie2009elements}. A common parametrization is through the effective degrees of freedom statistic df$_\l$.  We use the \texttt{R} package ``fields'' \cite{fieldsManual} to fit multi-dimensional thin plate splines, and the base \texttt{smooth.spline} function for the one-dimensional case.

\subsection{Kriging Surrogates}
A kriging  surrogate assumes that $f$ in \eqref{eq:oracle} has the form
\begin{equation}\label{eq:kriging}
f(z) = \mu(z)+X(z),
\end{equation}
where $\mu : \R^d \rightarrow \R$ is a trend function, and $X$ is a mean-zero square-integrable process. Specifically, $X$ is assumed to be a realization of a Gaussian process with covariance kernel $C$. The role of $C$ is identical to the regularized regression above, i.e.~$C$ generates the approximating family $\HH_C$ that $X$ is assumed to belong to.

However, kriging also brings a Bayesian perspective, treating $X$ as a random function to be learned, and estimation as computing the posterior distribution of $X$ given the collected data $\mathbf{y} \doteq (y^{(1)}, \ldots, y^{(N_{tr})})$. The RKHS framework implies that the posterior mean (more precisely its maximum a posteriori  estimate) of $X(z)$ coincides with the regularized regression prediction from the previous section. In the Bayesian framework, $C$ is interpreted as the covariance kernel, $C(z,z') = Cov( f(z), f(z'))$ as $f(\cdot)$ ranges over $\HH_C$. Assuming that the noise $\eps(z)$ is also Gaussian implies that $X(z) | \mathbf{y} \sim N( m(z), s^2(z))$ has a Gaussian posterior, which reduces to computing the kriging mean $m(z)$ and kriging variance $s^2(z)$.

In turn, the kriging variance $s^2(z)$ offers a principled empirical estimate of model accuracy, quantifying the approximation quality. In particular, one can use $s^2(z)$ as the proxy for the MSE of $\hat{f}$ at $z$. Integrating $s^2(z^{(n)})$ over the outer design locations then yields an assessment regarding the error of \eqref{eq:expected-F}.


\subsubsection{Simple Kriging} \label{sec:SimpleKriging}

Simple kriging (SK) assumes that the trend $\mu(z)$ is known.  By considering the process $f(z)-\mu(z),$ we may assume without loss of generality that $f(z)$ is centered at zero and $\mu \equiv 0$. The resulting posterior mean and variance are then \cite{roustant2012dicekriging}
\begin{equation}
\left\{ \begin{aligned}
m_{SK}(z) &\doteq \mathbf{c}(z)^T \mathbf{C}^{-1}\mathbf{y};\\
s^2_{SK}(z) &\doteq C(z,z)-\mathbf{c}(z)^T \mathbf{C}^{-1}\mathbf{c}(z),
\end{aligned} \right.
\end{equation}
where $\mathbf{c}(z) = \left(C(z,z^{(n)})\right)_{1 \leq n \leq N_{tr}}$ and
\begin{equation}
\mathbf{C} \doteq \left[C(z^{(i)},z^{(j)})\right]_{1 \leq i, j \leq N_{tr}}+\bm{\Delta},
\end{equation} with $\bm{\Delta}$ the diagonal matrix with entries $\tau^2(z^{(1)}), \ldots, \tau^2(z^{(N_{tr})})$.





\subsubsection{Universal Kriging}
Universal kriging (UK) generalizes \eqref{eq:kriging} to the case of a parametric trend function of the form $\mu(z) = \beta_0  + \sum_{j=1}^p \beta_j h_j(z)$ where $\beta_j$ are constants to be estimated, and $h_j(\cdot)$ are given basis functions. The coefficient vector $\bm{\beta} = (\beta_1, \dots, \beta_p)^T$ is estimated simultaneously with the Gaussian process component $X(z)$.  A common choice is first-order UK that uses $h_j(z) = z_j$ for $j=1,\ldots, d$. Another common choice is zero-order UK, also known as Ordinary Kriging (OK) that takes $\mu(z) = \beta_0$ a constant to be estimated.

If we let $\mathbf{h}(z) \doteq \left(h_1(z), \ldots, h_p(z)\right)$ and $\mathbf{H} \doteq \left(\mathbf{h}(z^{(1)}), \ldots, \mathbf{h}(z^{(N)})\right),$ then the universal kriging mean and variance at location $z$ are \cite{roustant2012dicekriging}
\begin{align}
\left\{ \begin{aligned}
m_{UK}(z) &= \mathbf{h}(z)^T \hat{\bm{\beta}} + \mathbf{c}(z)^T \mathbf{C}^{-1}(\mathbf{y}-\mathbf{H}\hat{\bm{\beta}});\\
s^2_{UK}(z) &= s^2_{SK}(z) + \left(\mathbf{h}(z)^T - \mathbf{c}(z)^T \mathbf{C}^{-1}\mathbf{H}\right)^T\left(\mathbf{H}^T \mathbf{C}^{-1} \mathbf{H}\right)^{-1} \left(\mathbf{h}(z)^T - \mathbf{c}(z)^T \mathbf{C}^{-1} \mathbf{H}\right),
\end{aligned}\right.
\end{align}
where the best linear estimator of the trend coefficients $\bm{\beta}$ is given by the usual linear regression formula $\hat{\bm{\beta}} \doteq \left(\mathbf{H}^T \mathbf{C}^{-1}\mathbf{H}\right)^{-1}\mathbf{H}^T \mathbf{C}^{-1}\mathbf{y}.$

The combination of trend and Gaussian process (GP) model offers an attractive framework for fitting a response surface. The trend  component allows to incorporate domain knowledge about the response, while the GP component offers a flexible nonparametric correction. One strategy is to specify a known trend (coming from some analytic approximation) and fit a GP to the residuals, yielding a Simple Kriging setup. Another strategy is to take a low-dimensional parametric approximation, such as a linear function of $Z$-components, and again fit a GP to the residuals, leading to a Universal Kriging setup.

\subsubsection{Covariance kernels and parameter estimation}
The covariance function $C(\cdot, \cdot)$ is a crucial part of a Kriging model.  In practice, one usually considers spatially stationary or isotropic kernels,  $$C(z,z') \equiv c(z-z') =  \s^2 \prod_{j=1}^d  g( (z-z')_j; {\theta}_j),$$
 reducing to the one-dimensional base kernel $g$. Below we use the power exponential kernels $g(h; \theta) = \exp\left(-\left(\frac{|h|}{\theta}\right)^p\right)$.  The hyper-parameters ${\theta}_j$ are called characteristic length-scales and can be informally viewed as roughly the distance you move in the input space before the response function can change significantly, \citet[Ch 2]{rasmussen2006gaussian}. The user-specified power $p \in [1,2]$ is usually taken to be either $p=1$ (the exponential kernel) or $p=2$ (the Gaussian kernel).
 Fitting a kriging model requires picking a kernel family and the hyper-parameters $\s_j, \theta_j$. Two common estimation methods are maximum likelihood, using the likelihood function based on the distributions described above, and penalized MLE (PMLE). 
 Either case leads to a nonlinear optimization problem to fit $\theta_j$ and process variance $\s^2$. One can also consider Bayesian Kriging, where trend and/or covariance parameters have a prior distribution, see \citet{helbert2009assessment}. We utilize the \texttt{R} package ``DiceKriging'' \cite{roustant2012dicekriging}
 that allows fitting of SK and UK models with five options for a covariance kernel family, and several options on how the hyper-parameters are to be estimated.




\subsubsection{Batching}
To construct an accurate emulator for $f(\cdot)$, it is important to have a good estimate of the sampling noise $\tau^2(z)$. Typically this information is not available to the modeler a priori. One of the advantages of plain nested Monte Carlo is that generating $N_{in}$ scenarios from a fixed $z^{(n)}$ gives natural empirical estimates \emph{both} for $f(z^{(n)})$ and $\tau^2(z^{(n)})$. To mimic this feature, we therefore consider batched or replicated designs $\mathcal{D}$. To wit, given a total budget of $N_{tr} = N_{tr,1} \cdot N_{tr,2}$ training samples, we allocate them into $N_{tr,1}$ distinct design sites $z^{(1)}, \ldots, z^{(N_{tr,1})}$, and then generate $N_{tr,2}$ trajectories from each $z^{(n)}$. Next, the above batches are aggregated into
\begin{align} \label{eq:batching}
  y^{(n)} &\doteq \frac{1}{N_{tr,2}} \sum_{j=1}^{N_{tr,2}} F(T, z^{(n),j}(\cdot) ); \\
  \hat{\tau}^{2}(z^{(n)}) &\doteq \frac{1}{N_{tr,2}-1} \sum_{j=1}^{N_{tr,2}} \left\{y^{(n)} - F(T, z^{(n),j}(\cdot) ) \right\}^2,
\end{align}
and the resulting dataset $\{ z^{(n)}, y^{(n)}, \hat{\tau}^2(z^{(n)})\}$, $n=1,\ldots, N_{tr,1}$ is used to fit a kriging model for $\hat{f}$, with $\hat{\tau}^2(z^{(n)})/N_{tr,2}$ proxying the simulation variance at $z^{(n)}$.

The efficient allocation between $N_{tr,1}$ and $N_{tr,2}$ was analyzed in \citet{broadie2011efficient} for a related risk management problem and it was shown that the optimal choices satisfy
\begin{equation}\label{eq:TrainingBudget}
N_{tr,1} \propto N_{tr}^{2/3}, \qquad N_{tr,2} \propto N_{tr}^{1/3}.
\end{equation}
This is also the allocation we pursue in this paper, so that there are relatively many more design sites than replications in each batch.


\subsection{Experimental Design} \label{sec:ChoiceOfDesign}
Several approaches are possible for constructing the training design $\mathcal{D}$. First, one may generate an empirical design by independently sampling $z^{(n)} \sim Z(T) | Z(0)$. This allows to emulate the conditional density $p_T(z| Z(0))$ which is advantageous for computing an expectation like in \eqref{eq:expected-F}. Second, one may generate a random $\mathcal{D}$ using some other proposal density $z^{(n)} \sim Q$. For example, a uniform proposal density (i.e.~$z^{(n)}$ i.i.d.~uniform in some domain $D \subseteq \R^d$) yields a basic attempt in having a space filling experimental design of arbitrary size. A more structured (but still random) design can be obtained via Latin Hypercube Sampling (LHS) techniques \cite{wyss1998user}.  Roughly speaking, LHS builds a regular $d$-dimensional lattice and then attempts to equidistribute $N_{tr,1}$ sites among the resulting hypercubes. Within each selected hypercube the design site is placed uniformly.


 Third, one can use a deterministic design, such as a latticed grid, or a quasi-Monte Carlo (QMC) sequence. Deterministic designs ensure a space-filling property and easy reproducibility.  For example, the Sobol  sequence  \cite{sobol1998quasi} redistributes a uniform binary grid to produce a grid that is maximally equidistributed. Compared to LHS, use of QMC is faster (as it can be directly hard-coded) and can be manually tweaked as needed. Both methods reduce Monte Carlo variance of $\hat{f}$ relative to empirical $\mathcal{D}$. Theoretically, the typical domain of $Z(T)$ is unbounded, e.g.~$\mathbb{R}+^d$. This is not an issue for empirical design construction; for LHS and QMC methods, one must specify an appropriate bounding domain $D \in \mathbb{R}^d$ before generating $\mathcal{D}$.

\begin{remark}
Depending on the context, the design $\mathcal{D}$ might need to be spatially non-uniform. For example, if using a deterministic design for computing \eqref{eq:expected-F}, it may be preferable to capture the correlation structure among the components of $Z(T)$, or to up-weigh the regions most likely for $Z(T)$. If one is estimating a quantile or tail expectation, $\mathcal{D}$ should preferentially cover the extreme values of the distribution of $Z(T)$; in that situation, an empirical design would be inappropriate.  
\end{remark}

\subsubsection{Generating Longevity Scenarios}
Construction of an emulator entails the basic building block of generating a longevity scenario $\{Z(t), t=0,\ldots\}$. In the simplest setting, this just requires to generate and manipulate a sequence of i.i.d~Uniform draws that describe the random increments of the (components) of $Z$. However, typically the model used also includes parameters that must be estimated or calibrated. This aspect becomes nontrivial when future longevity projections are made, whereby model re-fitting may be carried out. Re-fitting introduces path-dependency,
making parameters dynamic quantities that might need to be included in $Z$. For example, \citet{cairns2014longevity} advocate the PPC (partial parameter certain) scenario generation that breaks the overall simulation  into two pieces of $[0,T]$ and $[T,\infty)$. With PPC, one initially calibrates the model at $t=0$ using past mortality data and then simulates up to time $T$. The simulated scenario is then appended to the historical data, so that the simulation becomes the new ``history'' from time 0 to time $T.$ The model parameters are then re-fitted at $T$  and the resulting, modified longevity dynamics of $Z$ are used to simulate beyond $T$. The idea of PPC is to capture some memory of mortality evolution, in essence removing some of the presumed Markovian structure. Under PPC the refitted parameters are blended into $Z(T)$ since they affect the resulting $F(T, Z(\cdot))$.

Also, in the interest of dimension reduction, one could drop some components of the full state space when constructing the emulator. To do so, one may analyze what dynamic variables materially impact annuity values, for example via some simple regression models to test for statistical significance.

\subsection{Fitting and Evaluation of Emulators} \label{sec:fitting}

To fit an emulator for a given simulation budget $N_{tr}$, we first decompose $N_{tr} = N_{tr,1} \times N_{tr,2}$ and then construct an experimental design $\mathcal{D}$ of size $N_{tr,1}$ using one of the methods in Section \ref{sec:ChoiceOfDesign}. Each site in $\mathcal{D}$ then spawns $N_{tr,2}$ trajectories that are batched together as in \eqref{eq:batching}.
%
Fitting is done in \texttt{R} using the mentioned publicly available packages. For kriging, we use the default setting of the \texttt{km} function in DiceKriging package.

Given $\hat{a}(z,T,x)$ we evaluate its performance across a test set $\mathcal{D}^{test} = (z^{(1)}, \ldots, z^{(N_{out})})$ of $N_{out}$ locations. Note that $\mathcal{D}^{test}$ is distinct from the training set $\mathcal{D}$. In line with \eqref{eq:expected-F} we use an empirical testing set $\mathcal{D}^{test}$: $z^{(n)} \sim Z(T)|Z(0)$. Since the true values $a(z,T,x)$ are not available, we benchmark against an (expensive) gold standard estimate $\hat{a}^{MC}(z,T,x)$ that is described below. In particular, we record the integrated MSE and Bias statistics from \eqref{eq:IMSEBias}, namely
\begin{align}
\label{eq:IMSEMC}
\widehat{\text{IMSE}}\left(\hat{a}\right) &= \frac{1}{N_{out}}\sum_{n=1}^{N_{out}} \left(\hat{a}(z^{(n)},T,x) - \hat{a}^{MC}(z^{(n)},T,x)\right)^2;\\ \label{eq:bias}
\widehat{\text{Bias}}\left(\hat{a}\right) &= \frac{1}{N_{out}}\sum_{n=1}^{N_{out}} \left[ \hat{a}(z^{(n)},T,x) - \hat{a}^{MC}(z^{(n)},T,x) \right].
\end{align}

For benchmarking, we use a high-fidelity nested Monte Carlo approach \eqref{eq:N-out}-\eqref{eq:N-in}. While expensive, it is a simple, asymptotically consistent, unbiased estimator. Specifically, for valuing annuities, $\hat{a}^{MC}(z,T,x)$ is obtained by averaging $N_{in} = 1000$ scenarios of $\{Z(s), s > T\}$ at each $z^{(n)} \in \mathcal{D}^{test}$. Unless indicated otherwise, we use $N_{out} = 1000$, so that the overall budget of $\hat{a}^{MC}$ is $\mathcal{O}(N_{out} \times N_{out})$. We then compare against emulators that use $N_{tr} \in [100, 8000]$, which yields an efficiency gain on the order of 10-50x speed-up. We also compare against  deterministic estimators that  require no training at all (but do need an analytic derivation), and take just $\mathcal{O}(N_{out})$ budget to make predictions for the outer $N_{out}$ simulations to evaluate \eqref{eq:IMSEMC}.

\section{Case Study: Predicting Annuity Values under a Lee-Carter with Shocks Framework} \label{sec:chen-coxcasestudy}

\citet{chen2009modeling} introduced a mortality model based on the traditional Lee Carter set-up:
\begin{equation}
\log m(t,x) = \beta^{(1)}(x) + \beta^{(2)}(x) \kappa^{(2)}(t). \label{eq:LeeCarter}
\end{equation}
This is the same as the APC model (M2) in \ref{sec:leecarter-cbd} without the cohort term. In the Chen-Cox model, $\beta^{(1)}(x)$ and $\beta^{(2)}(x)$ are deterministic vectors capturing age effects, and $\kappa^{(2)}(t)$ is a stochastic process capturing the period effect with dynamics
\begin{equation}
\label{eq:kappaLCjumps} \kappa^{(2)}(t+1)=\kappa^{(2)}(t) + \mu^{(1)} + \xi^{(1)}(t+1) + [\xi^{(2)}(t+1) - \xi^{(2)}(t)],
\end{equation}
where $\xi^{(1)}(t) \sim N(0, \s^{(1)})$ and $\xi^{(2)}(t)$ has an independent zero-modified normal distribution with $\P(\xi^{(2)}(t) = 0) = 1-p,$ and Gaussian parameters $(\mu^{(2)},\s^{(2)}).$  The motivation for \eqref{eq:kappaLCjumps} is to incorporate idiosyncratic mortality shocks represented by $\xi^{(2)}$, that occur with probability $p$ any given year and have a random magnitude with distribution $N(\mu^{(2)}, \s^{(2)})$. Such shocks, representing natural or geopolitical catastrophes, are temporary and last just a single period, hence subtraction of the last term $-\xi^{(2)}(t)$ in \eqref{eq:kappaLCjumps}. Due to this term, it would appear that the model has a two-dimensional state space $\{\kappa^{(2)}(t), \xi^{(2)}(t)\}$. However, we note that it is sufficient to generate scenarios starting with $\kappa^{(2)}(T)$ and assuming $\xi^{(2)}(T) = 0$ (no shock in year $T$). Then after estimating $f(\kappa) = \E[ F(T,\kappa^{(2)}(\cdot) ) | \kappa^{(2)}(T) = \kappa, \xi^{(2)}(T) = 0]$, one easily obtains in case of year-T shocks $\E[ F(T,\kappa^{(2)}(\cdot) | \kappa^{(2)}(T) = \kappa, \xi^{(2)}(T) = \xi] = f(\kappa- \xi)$, reducing to the prediction of ``unshocked'' values.

The presence of idiosyncratic shocks in $m(t,x)$ renders the corresponding survival probability analytically intractable. However, the linear dynamics of $\kappa^{(2)}$ in \eqref{eq:kappaLCjumps} allows to obtain the following deterministic estimator for future mortality rates.

\begin{lemma}\label{lemma:chen-cox}
Let $Z(s) = \{ \kappa^{(2)}(s), \xi^{(2)}(s) \}$. Under the Chen-Cox model, the following holds:
\begin{equation}
\E[\kappa^{(2)}(t) \mid Z(s)] = \kappa^{(2)}(s) + (t-s)\mu^{(1)} + \mu^{(2)}p - \xi^{(2)}(s), \qquad 0 \leq s \leq t < \infty. \label{eq:chencoxexpectation}
\end{equation}
\end{lemma}

The proof can be found in \ref{sec:proofs}.  Substituting \eqref{eq:chencoxexpectation} into \eqref{eq:LeeCarter} yields the following estimator for $\E[m(T+s,x) \mid \kappa^{(2)}(T), \xi^{(2)}(T)]:$
\begin{equation}\label{eq:mhatchen-cox}
\hat{m}(T+s,x)\doteq\exp\left(\beta^{(1)}(x)+\beta^{(2)}(x)\left(\kappa^{(2)}(T) + s\mu^{(1)} + \mu^{(2)}p - \xi^{(2)}(T) \right)\right).
\end{equation}

%

\subsection{Results}\label{sec:chencoxresults}

We follow \citet{chen2009modeling} in using US mortality data obtained from the National Center for Health Statistics (NCHS)\footnote{Source: \url{http://www.cdc.gov/nchs/nvss/mortality_tables.htm}}. This dataset contains yearly age specific death rates for overall US population over 1900--2003. Fitting  yields the random-walk parameters $\mu^{(1)} = -0.2173, \s^{(1)} = 0.3733$ in \eqref{eq:kappaLCjumps}, as well as the estimated probability of shock as $p=0.0436$, with jump distribution $(\mu^{(2)},\s^{(2)}) = (0.8393, 1.4316)$. As expected, $\mu^{(2)} \gg 0$ is large and positive, so shocks correspond to large temporary increases in mortality.
The goal is to analyze and compare the ability of kriging models and analytic estimates to predict  $T=10$-year deferred annuity values for unisex $x=65$ year olds. Payments are cut-off at age $\bar{x}=94$. We use a discount rate of $r=4\%$.



\begin{table}[ht]
\begin{center}
\begin{tabular}{r l l  l l  ll}
\hline
 & \multicolumn{2}{c}{$N_{tr}=125$} & \multicolumn{2}{c}{$N_{tr}=512$}& \multicolumn{2}{c}{$N_{tr}=1000$}\\
Type & \multicolumn{1}{c}{Bias} & \multicolumn{1}{c}{$\sqrt{\text{IMSE}}$} &  \multicolumn{1}{c}{Bias} & \multicolumn{1}{c}{$\sqrt{\text{IMSE}}$} & \multicolumn{1}{c}{Bias} & \multicolumn{1}{c}{$\sqrt{\text{IMSE}}$}\\ \hline
Analytic &   1.668e-03  &  2.148e-03  & 1.668e-03  &  2.148e-03  & 1.668e-03  &  2.148e-03 \\
Ord.~Kriging &   5.145e-03  &  5.923e-03  & 1.582e-04  &  1.975e-03    &  -1.999e-04  &  1.634e-03   \\
Univ.~Kriging &    5.832e-03  &  6.059e-03  & 4.816e-04  &  1.045e-03  &   -1.243e-05  &  7.428e-04    \\  \hline
\end{tabular}
\end{center}
\caption{Comparing estimators for life annuity value under the Chen-Cox model for different size of experimental design.  The design $\mathcal{D}$ is constructed with $N_{tr}=N_{tr,1}^{2/3}\cdot N_{tr,2}^{1/3}.$  The reported values are evaluated from a Monte Carlo benchmark, using \eqref{eq:IMSEMC} and \eqref{eq:bias}. Analytic estimate is based on \eqref{eq:mhatchen-cox}; universal kriging model uses first-order linear basis functions.}
\label{tab:chencoxcomparison}
\end{table}

W fit emulators with budgets $N_{tr} \in \{125, 512, 1000\}$. The respective training designs $\mathcal{D}$ are deterministic and uniformly spaced across an appropriately chosen interval $D = [ \underline{\kappa}, \bar{\kappa}]$; a fixed design minimizes Monte Carlo noise in fitting $\hat{f}(\cdot)$.  Because $Z \equiv \kappa^{(2)}$ is just one-dimensional, a relatively small training budget is used. For the emulators, we fit both an ordinary kriging (OK) model with constant trend $\mu(\kappa) = \beta_0$, and first-order linear universal kriging (UK) model with $\mu(\kappa) = \beta_0 +  \beta_1\kappa$. For evaluation, we fix a testing set containing $N_{out}=50$ values of $Z(T),$ each with a Monte Carlo benchmark containing $N_{in}=10^5$ simulations. Due to the very small MSE's involved, a very high-fidelity benchmark was needed (in order to isolate the MSE of the emulator from the MSE of the benchmark), leading to a very large $N_{in}$. To be computationally feasible, we picked a small testing set. To make sure that $\mathcal{D}^{test}$ accurately represents the distribution of $Z(T)$ its locations were picked as the empirical $1\%, 3\%, \ldots, 99\%$ percentiles of a large sample of $Z(T)$. The mortality shocks associated with these percentiles were used in the comparison process.

\begin{figure}
\begin{center}
\includegraphics[scale=0.4]{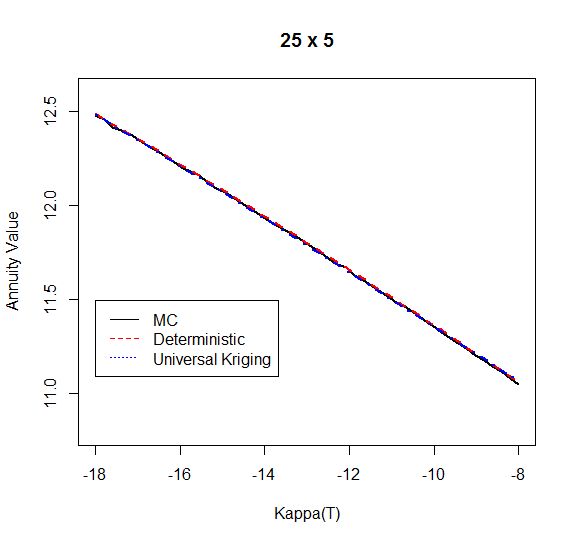}
\includegraphics[scale=0.4]{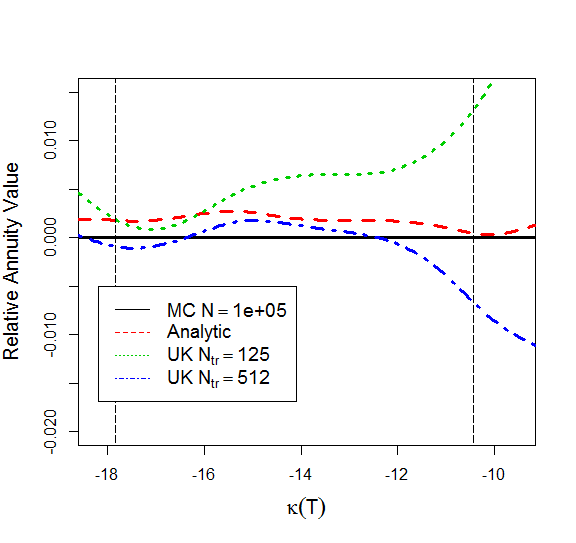}
\end{center}
\caption{Annuity emulators in the Chen-Cox model. Left: three estimators (MC, UK w/$N_{tr}=125$ and analytic) of annuity value $a(\kappa^{(2)}(T))$ vs.~$\kappa^{(2)}(T)$.  The training design (indicated by the vertical dashed lines) is $\mathcal{D} = \{\kappa^{(2)}(T) \in (-17.5,-10)\}$ with $N_{tr,1} = 25, N_{tr,2} = 5$. Right: relative annuity values vis-a-vis the Monte Carlo
 benchmark $\hat{a}^{MC}$ obtained with $N_{in} = 10^5$. }
\label{fig:chenline}
\end{figure}

Table \ref{tab:chencoxcomparison} and Figure \ref{fig:chenline} summarize the results. We observe that there is quite a wide spread in potential future annuity prices, with differences of more than 10\% (or \$1 in annuity NPV) depending on realized $Z(T)$. This confirms the significant level of longevity risk. As shown in the Figure \ref{fig:chenline}, there is a nearly linear relationship for $z \mapsto a(z,T,x)$, which is perhaps surprising given the above range of forecasts. This strong linear trend in the response partly explains the advantage of the UK model over OK.
  The Figure also reflects the effect of training set size and distribution: the $N_{tr}=512$ model performs significantly better than its $N_{tr}=125$ counterpart. We see that all methods perform well, with IMSE's on the order of 1e-03. Even though the computed biases are rather small, we remark that since pension portfolios have very large face values, the corresponding approximation errors could be financially meaningful. For example, for a modest pension fund with an obligation of \$100mm, a bias of 1e-03 implies inaccuracy of \$100k.

The right panel of Figure \ref{fig:chenline} provides a zoomed-in visualization of the estimators' bias relative to $\hat{a}^{MC}$. As expected, the analytic estimator based on Lemma \ref{lemma:chen-cox} overestimates the true annuity value for all $\kappa^{(2)}(T)$. For $N_{tr}=125$, the kriging emulator clearly has a larger MSE, and in this case typically under-estimates $a(\kappa^{(2)}(T))$. For $N_{tr}=512$ we observe the statistical learning taking place, as the kriging model now has an excellent fit in the middle of the plot and essentially zero bias averaging over potential values of $\kappa^{(2)}(T)$. The effect of larger training budget is confirmed in Table \ref{tab:chencoxcomparison}, with IMSE's all decreasing towards zero as $N_{tr}$ increases.

The above analysis demonstrates that in some settings, the shape $z \mapsto a(z,T,x)$ is sufficiently simple that little modeling is required, and analytic estimators perform well (as do statistical emulators). However, we stress that there is no easy way to tell a priori that the analytic estimator would be adequate, and in any case a sufficiently large training set size will guarantee a better predictive power for the kriging models.

\begin{figure}
\begin{center}
\includegraphics[scale=0.4]{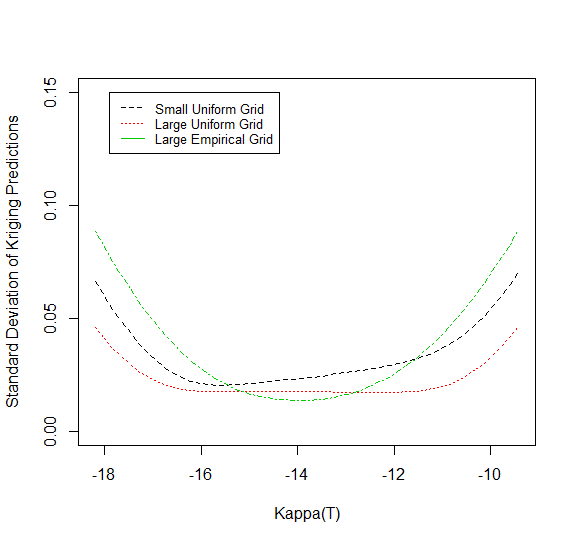}
\end{center}
\caption{Effect of training design $\mathcal{D}$ on the emulator accuracy in the Chen-Cox model. We show kriging standard deviation $s(z)$ for the universal kriging model with three different designs: $\mathcal{D}^{(A)}$ (small uniform), $\mathcal{D}^{(B)}$ (large uniform) and $\mathcal{D}^{(C)}$ (large empirical). The deterministic designs $\mathcal{D}^{(A)},\mathcal{D}^{(B)}$ contain uniformly spaced values of $\kappa^{(2)}(T) \in (-17.5, 10),$ of size $N_{tr}^{(A)} = 125$ and $N_{tr}^{(B)}=1000$ respectively.  $\mathcal{D}^{(C)}$ is an empirical design of size $N_{tr}^{(C)} = 1000$ generated using the density of $\kappa^{(2)}(T) | \kappa^{(2)}(0)$.}
\label{fig:chencoxgrideffect}
\end{figure}

 The one dimensional case also provides a visual representation of the effect of grid design, illustrated in Figure~\ref{fig:chencoxgrideffect}.  The figure showcases two features of emulators: (i) dependence between local accuracy as measured by $s^2(z)$ and grid \emph{size} $N_{tr}$; and (ii) dependence between $s^2(z)$ and grid \emph{shape}.
 First, larger training sets improve accuracy (with a general relationship of $\mathcal{O}( N_{tr}^{-1/2})$ like in plain Monte Carlo). This can be seen in Figure~\ref{fig:chencoxgrideffect} where kriging standard deviation $s(z)$ is consistently lower for $N^{(B)}_{tr}=1000$ compared to $N^{(A)}_{tr} = 125$. One implication is that as $N_{tr} \to \infty$, we would have $s^2(z) \to 0$, i.e.~$f(\cdot)$ would be learned with complete precision, a property known as global consistency of the emulator. Second, $s^2(z)$ is affected by the shape of $\mathcal{D}$ in the sense that higher local density of training points lowers the local posterior variance. This is intuitive if viewing $\hat{f}$ as an interpolator or kernel regressor -- the denser the training set around $z$, the better we are able to infer $f(z)$. Consequently, the empirical grid $\mathcal{D}^{(C)}$ that is concentrated around the mode of $Z(T)$, offers better accuracy in that neighborhood (around $\kappa^{(2)}(T) \simeq -14$ in Figure~\ref{fig:chencoxgrideffect}) compared to the uniform $\mathcal{D}^{(B)}$, but lower accuracy towards the edges, where $\mathcal{D}^{(C)}$ becomes sparser. For all designs, posterior uncertainly deteriorates markedly as we migrate outside of the training set (e.g.~$\kappa^{(2)}(T) > -11$ in the Figure).

The $s(z)$ values shown in Figure \ref{fig:chencoxgrideffect} also provide an approximation of emulator IMSE. For example, averaging the kriging standard deviation $s(z)$ over the testing set using the UK model with $N_{tr}=1000$ yields $s_{Ave} = \sqrt{ \{ \frac{1}{N_{out}} \sum_n s^2(z^{(n)}) \} } = 7.159e-03$, while in Table \ref{tab:chencoxcomparison} the corresponding reported IMSE was $\widehat{IMSE} = 7.428e-04$. Reasons for the mismatch include the residual MSE in the Monte Carlo estimate $\hat{a}^{MC}$ and model mis-specification of the UK model, which would bias the self-assessed accuracy. Moreover, the strong correlation between $\hat{a}(z)$ across different testing locations $z^{(n)}$ implies that $\widehat{IMSE}$ has a large standard error. Nevertheless, $s_{Ave}$ is a highly useful metric that allows to quantify the relative accuracy of different emulators in the absence of any gold-standard benchmarks.


 %
%

\section{Case Study: Hedging an Index-Based Fund in a Two-Population Model} \label{sec:2pop-study}


There has been a lot of recent discussion regarding index-based longevity funds.  Information on the death rates of the general public is widely available, and a market fund that uses the respective death rats as its price index offers a standardized way to measure population longevity. In particular, it allows for securitization of longevity swaps that can be used by pension funds to hedge their longevity risk exposure. If the pension fund could buy as many units of the swap as it has to pay out to its annuitants, it would result in a situation where the amount paid is nearly equal to the amount received from the swap. The quality of such as hedge is driven by the basis risk between the indexed population and the annuitant pool, that is typically a subset of the index. Consequently, it is necessary to create a model to capture the link between the index and the insured sub-population.

\begin{remark}
From a different angle, some longevity products explicitly integrate mortality experience in several regions, for example across different countries (UK, Germany, Netherlands) or across different constituencies (England vis-a-vis Great Britain).  \citet{lin2013pricing} states that most mortality data reported by official agencies calculate a weighted average mortality index of different underlying populations.  They also investigate the modeling aspect of such multi-population indices.
\end{remark}

To fix ideas, we call the index population Pool 1, and the annuitants Pool 2.
Consider now an individual from Pool 2 who will be aged $x$ at date $T$ when she begin to receive her life annuity. The corresponding time-$T$ liability to the pension fund is denoted $a_2(Z(T),T,x)$. If the pension fund enters into a swap based on the index, she might purchase $\pi$ index-fund annuities for age $x$, with net present value of $\pi a_1(Z(T),T,x),$ at $T$. For now we ignore what would be a fixed premium.  The overall hedge portfolio is then $\Delta(Z(T),T,x) \doteq \pi a_1(Z(T),T,x)-a_2(Z(T),T,x)$. Several risk measures can be used to determine hedge effectiveness.  Some examples include variance, or tail risk measures such as value-at-risk (VaR) or expected shortfall (TVaR). Recent work in this direction includes \citet{coughlan2011longevity} who used a bootstrapping and extrapolation method to analyze hedge effectiveness, and \citet{cairns2014longevity} whose setup we follow below.

Unsurprisingly, the correlation structure for mortality across populations is complex. One notable recent contribution is by \citet{cairns2011bayesian,cairns2014longevity} who considered a hedging problem between an index pool $k=1$ and insured sub-pool $k=2$.  Specifically, the two populations are the England \& Wales (E\&W) general population, which represents the index mortality rate (Pool 1), and the Continuous Mortality Investigation (CMI) population, which are mortality rates gathered from United Kingdom insured populations, serving the role of those receiving pension payments (Pool 2). To model the dependence between the two pools, \citet{cairns2011bayesian} proposed a cointegrated two-population Bayesian model based on the Lee-Carter framework. To wit, the mortality rates $m_k(t,x)$ behave similar to \eqref{eq:m2},
\begin{align*}
  \log m_k(t,x) &= \beta^{(1)}_k(x) + n_a^{-1} \kappa^{(2)}_k(t) + \gamma^{(3)}_k(t-x), \quad k=1,2
\end{align*}
with stochastic dynamics of the period effect $\kappa^{(1)}_1$ given by

\begin{align} \kappa^{(2)}_1(t) = \kappa^{(2)}_1(t-1) + \mu_1 + \s_1 \epsilon_1(t), \quad \epsilon_1(t) \overset{\text{i.i.d}}{\sim} N(0,1).\end{align}
In turn, the mortality of the larger population influences the period effect of the smaller (insured) population, with dynamics for $\kappa^{(2)}_2$ co-integrated with $\kappa^{(1)}_1$. Namely, their difference $S(t) \doteq \kappa^{(2)}_1(t) - \kappa^{(2)}_2(t)$ forms an $AR(1)$ process
\begin{align}\label{eq:S(t)}
S(t) &= \mu_2 + \phi(S(t-1)-\mu_2) + \s_2 \epsilon_2(t-1) +
c \epsilon_1(t-1), \quad \epsilon_2(t) \overset{\text{iid}}{\sim} N(0,1),\end{align}
with $\epsilon_1(\cdot)$ being independent of $\epsilon_2(\cdot)$, and $c = \s_1 - \rho \s_2$ for the covariance, where $\rho = \text{Corr}(\kappa^{(2)}_1(t), \kappa^{(2)}_2(t))$.  In both models cohort effects $\gamma^{(3)}_k$ are independent $AR(2)$ processes.  Here, $\phi$ is the mean-reversion rate.  Since \eqref{eq:S(t)} models the difference $\kappa^{(2)}_1(t)-\kappa^{(2)}_2(t),$ $\phi$ reflects the rapidity of how $S(t)$ returns to the constant $\mu_2$, which is assumed to be the constant difference between the two populations.


\subsection{Analytic Approximations}\label{sec:analytic2pop}

\citet{cairns2014longevity} used the fact that $\E[\kappa^{(2)}_1(T+t) \mid \kappa^{(2)}_1(T)] = \kappa^{(2)}_1(T) + \mu_1 t$ to introduce the median-mortality approximation
\begin{equation}\label{eq:mMM1}
\hat{m}^{A1}_1(T+t,x+t) = \exp\left( \beta^{(1)}_1(x+t) + \frac{1}{n_a} (\kappa^{(2)}_1(T) + \mu_1 t) + \frac{1}{n_a} \gamma^{(3)}_1(T-x)\right).
\end{equation}
Since $S(t)$ is mean reverting, it is also suggested to use the approximation for the CMI population of
\begin{equation}\label{eq:cairnsMM2}
\hat{m}^{A1}_2(T+t,x+t) = \exp\left( \beta^{(1)}_2(x+t) + \frac{1}{n_a} (\kappa^{(2)}_2(T) + \mu_1 t) + \frac{1}{n_a} \gamma^{(3)}_2(T-x)\right),
 \end{equation}
i.e.~the same drift as the general population but different initial value.

We introduce a different, more accurate approximation based on the following lemma.
\begin{lemma}\label{lemma:two-pop} We have
\begin{align} \E\left[\kappa^{(2)}_2(T+t) | Z(t) \right] = \kappa^{(2)}_1(T) + \mu_1 t - \mu_2(1-\phi^t) - \phi^t(\kappa^{(2)}_1(T) - \kappa^{(2)}_2(T)).\end{align}
\end{lemma}
The proof can be found in \ref{sec:proofs}.  Denote $\E[\kappa^{(2)}_2(T+t) | Z(t)] \doteq \xi(t,T).$  Lemma \ref{lemma:two-pop} suggests an alternative analytic estimator for $m_2(T+s,x)$ as
\begin{equation}\label{eq:mMM2}
 \hat{m}^{A2}_2(T+s,x+s) \doteq \exp\left( \beta^{(1)}_2(x+s) + \xi(t,T) + \frac{1}{n_a} \gamma_2(T-x)\right).
\end{equation}


Denote $a_1(Z(T))$ and $a_2(Z(T))$ as the  net present value at $T$ (conditional on $Z(T)$) of a life annuity for the E\&W and CMI populations respectively as defined in \eqref{eq:life-annuity}.  In what follows, \textit{Analytic 1} will refer to use of \eqref{eq:mMM1} and \eqref{eq:cairnsMM2} in estimating survival probabilities \eqref{eq:P2} for each population (and hence $a_1$ and $a_2),$ while \textit{Analytic 2} refers to the use of \eqref{eq:mMM1} and \eqref{eq:mMM2}.  Notation for deferred annuity values under the two analytic approaches will be $\hat{a}_k^{A1}(z)$ and $\hat{a}_k^{A2}(z), k=1,2.$ 

\subsection{Model Fitting}
The parameters $\beta_1^{(1)}(x)$, $\beta_2^{(1)}(x)$, and past trajectories $\kappa^{(2)}_k(t)$, $\gamma^{(3)}_k(t-x)$, for $k=1,2$ were estimated from the male E\&W and CMI populations respectively, and the time and age ranging from calendar years 1961 to 2005 (with 2005 treated as $t=0$), and $x$ from $50$ to $89.$  The processes $(\kappa^{(2)}_1(t))$ and $(S(t))$ were fit as random walk with drift and $AR(1)$ respectively, introducing additional parameter estimates for $\mu_1, \s_1, \mu_2, \phi, \s_2$ and $c.$
We find $\mu_1 = -0.5504, \mu_2 = 0.6105, \s_1 = 1.278, \s_2 = 0.568, \phi = 0.9407, c=0.262$, so that the CMI population tends to have higher mortality, with a co-integration of about 94\%.

Using the PPC approach of \cite{cairns2014longevity},
we treat the age-effect parameters as fixed, and refit the $ARIMA$ models at period $T$ for each simulation.  That is, the $\beta^{(1)}_k$ are fixed throughout for $k = 1, 2$ and each of $\mu_1$, $\s_2$, $\mu_2$, $\phi,$ $\s_2$, and $c$ are re-estimated.  In principle, this makes the re-estimated parameters part of the state variable $Z(T).$  A few preliminary runs indicate that the variance parameters $\s_1, \s_2$ and $c$ have little significant effect on annuity values, while $\mu_1, \mu_2$ and $\phi$ do.  Since $\mu_1$ is in one-to-one correspondence with $\kappa^{(2)}_1(T),$ our time $T$ state process is finally characterized as $$Z(T) = \{\kappa^{(2)}_1(T), \kappa^{(2)}_2(T), \mu_2, \phi\}.$$  Heuristically, this is a reasonable choice: each element of $Z(T)$ has a direct effect on the  time $T$ mortality rates or their trends, while the variance terms simply add variability.

Several stochastic mortality models have \texttt{R} code available\footnote{LifeMetrics Open Source R code for Stochastic Mortality Modelling; see \url{http://www.macs.hw.ac.uk/~andrewc/lifemetrics/} for details} for model fitting.  We use the code to fit the two-population model parameters, yielding the inferred past trajectories for the age, period, and cohort effects. In a separate step, the estimated period and cohort effects are modeled as individual $ARIMA$ models.

For the remainder of this section we assume the starting age of the annuitant is $x=65$ with a fixed interest rate of $r=0.04$ and a $T=10$ year deferral period.  Generally the choice of hedge ratio $\pi$ is chosen systematically, for example through minimizing variance.  In this paper we assume the neutral value of $\pi=1$ in order to not favor one estimation type over another.  Hence the value of the hedge portfolio is $\Delta(Z(T)) = a_1(Z(T))-a_2(Z(T)).$


As discussed in Section \ref{sec:ChoiceOfDesign}, determining the training set design depends on the problem at hand.  In our particular example with a $4$-dim $Z(T)$, we aim to give an accurate result of the expectation of the hedge portfolio $\Delta(T)$, so we use an empirical design, as suggested in Section \ref{sec:ChoiceOfDesign}.  This also holds the advantage of capturing the correlation between $\kappa^{(1)}$ and $\kappa^{(2)}$ which is important in this co-integrated model.  To compare the effect of budget size, we choose two different budgets, $N_{tr}=1000$ and $N_{tr}=8000.$  Following the framework in Section \ref{sec:ChoiceOfDesign}, $N_{tr}$ is allocated into $N_{tr,1}=N_{tr}^{2/3}, N_{tr,2}=N_{tr}^{1/3},$ so that we have $N_{tr,1}=100$ (resp.~$N_{tr,1}=400$) training points with Monte Carlo simulations containing $N_{tr,2}=10$ (resp.~$N_{tr,2}=20$) batched simulations for each design point.

Different surrogate models are chosen than in Section \ref{sec:chen-coxcasestudy}; this time around a multi-dimensional state process suggests the use of a TPS model from Section \ref{sec:splines}.  We forego the OK model, but maintain use of the 1st-order linear UK model, and also implement a simple kriging (SK) model with $\mu(z)=\hat{a}_1^{A2}(z)-\hat{a}_2^{A2}(z).$  This combines advantages from both the analytic and UK approach, giving us an already accurate estimate for the trend, while nonparametrically modeling the residuals.  For these reasons, a SK emulator should outperform both the analytic estimators and the UK model.  We utilize another advantage of the surrogate models and fit them directly to the hedge portfolio values $\Delta(Z(T))$ rather than individually modeling annuity values $a_k(Z(T))$ and then taking difference of two approximations.

The Monte Carlo benchmark yields an average portfolio value of 0.1995 with a standard deviation of 0.1067.  This suggests that a one-to-one purchase of index annuity is not the optimal hedge unit under this population model.  In an actual application, one would analyze further to determine a different choice for $\pi$; for example \citet{cairns2014longevity} chooses $\pi$ to minimize portfolio variance.

\subsection{Results}\label{sec:2popresults}
\begin{table}[ht]
\begin{center}
\begin{tabular}{rcccc} \hline
& \multicolumn{2}{c}{$N_{tr}=1000$} & \multicolumn{2}{c}{$N_{tr}=8000$} \\
Type & Bias & $\sqrt{\text{IMSE}}$  & Bias & $\sqrt{\text{IMSE}}$ \\ \hline
Analytic A1 from \eqref{eq:cairnsMM2} & -2.101e-02  &  3.460e-02  & -2.101e-02  &  3.460e-02     \\
Analytic A2 from \eqref{eq:mMM2} & 4.480e-03  &  5.321e-03    & 4.480e-03  &  5.321e-03   \\
Thin Plate Spline & 2.577e-03  &  1.304e-02  & 5.803e-04  &  5.095e-03 \\
Universal Kriging &  4.363e-04  &  1.856e-02  &   1.857e-03  &  1.289e-02  \\
Simple Kriging &  -1.334e-03  &  3.280e-03 &  9.390e-04  &  3.043e-03    \\ \hline
\end{tabular}
\end{center}
\caption{Performance of analytic estimates and surrogate models for hedge portfolio values in the two-population model case study.  Numbers reported are based on $N_{out}=1000$ simulations of $Z(T)$ with a Monte Carlo benchmark.  $N_{tr}$ is allocated into $N_{tr,1} = N_{tr}^{2/3}$ training points and $N_{tr,2}=N_{tr}^{1/3}$ Monte Carlo batches per training point. Simple kriging model uses A2 estimator as trend. 
}
\label{tab:2popresults}
\end{table}

We choose $N_{out}=1000$ simulations of $Z(T)$ and predict hedge portfolio values $\Delta(Z(T))$ with the surrogate models, as well as via the deterministic estimates. Table \ref{tab:2popresults} shows the results.  As expected, the Analytic A2 estimator outperforms Analytic A1 since it is catered directly to the two-population model.  Relative to A1, our improved estimator cuts bias by nearly 80\%.
As for the surrogates, when $N_{tr}=1000,$ each of the TPS and UK models only slightly underperform the analytic estimate A2, while the SK model does significantly better.  For $N_{tr}=8000,$ both TPS and SK are better than A2.

\begin{figure}
\begin{center}
\includegraphics[scale=0.35,trim=0.7in 0.7in 0.6in 0.6in]{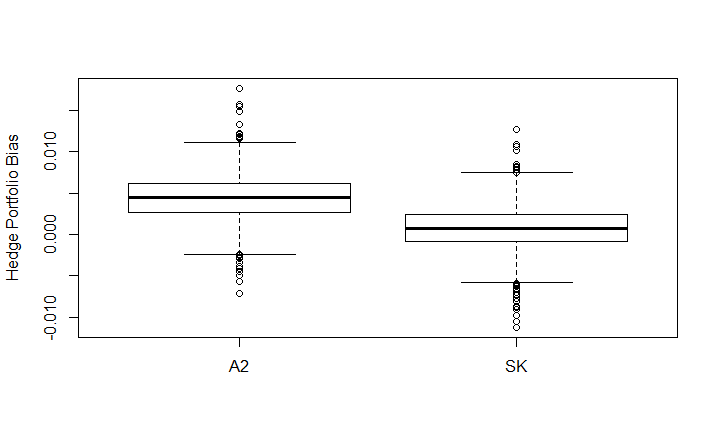}
\end{center}
\caption{Boxplots of hedge portfolio value bias for $N_{tr}=8000$ for analytic A2 and simple kriging approaches. To construct the boxplot, we computed for each of 1000 simulated values of $Z(T)$, the difference between the respective estimate and the Monte Carlo benchmark.}
\label{fig:boxplotbias}
\end{figure}

Figure \ref{fig:boxplotbias} summarizes the empirical distribution of the bias of the A2 and SK estimators given simulations of $Z(T)$.  We can see that both approaches have similar variability, while SK has a much lower bias.  The UK and TPS estimators have similar distributions with slightly larger bias than SK.

There are a few comments to be made in regards to these results.  First of all, there is no way to tell a priori that a deterministic estimate will perform well.  For example each surrogate model completely outclasses A1, while TPS and UK perform only marginally better than A2.  Possibly, even better (or worse) analytic estimators can be derived.  Additionally, the deterministic estimators are for annuity values themselves and not for the portfolio difference $\Delta(T)$.  A lower bias for a $\Delta(T)$ could simply be a consequence of the bias of each annuity $a_k(Z(T))$ being reduced during subtraction.

\section{Case Study: Predicting Annuity Values under the CBD Framework}\label{sec:cooked}

\subsection{Model Fitting}
 Our third case study utilizes another popular class of mortality models, the CBD \cite{cairns2006two} models which directly work with the survival probabilities $P$. To wit, we model the 1-period survival probability
\begin{equation}\label{eq:cbd}
P(Z(T);T,1,x)= \frac{1}{1+\exp\left(\kappa^{(1)}(T)+(x-{x}_{Ave})\kappa^{(2)}(T)\right)},
\end{equation}
where ${x}_{Ave} = n_a^{-1} \sum_i x_i$, and $\kappa^{(1)}, \kappa^{(2)}$ follow $ARIMA$ models, which according to \citet{cairns2011mortality} provide a good fit for period effects. Multi-period survival probabilities are obtained as products of \eqref{eq:cbd}.

We fit \eqref{eq:cbd} to the CMI population, considering a full range of $ARIMA(p,d,q)$ models with $p,q = 0,1,2,3,4$ and $d=0,1,2,$ using \texttt{auto.arima} in \texttt{R} from the package ``forecast'' \cite{hyndman2015forecast}.  The optimal configuration for this population is for $\kappa^{(1)}$ to follow $ARIMA(0,1,3)$ with drift and $\kappa^{(2)}$ to follow $ARIMA(1,1,2)$:
\begin{align}
\kappa^{(1)}(t) &= \kappa^{(1)}(t-1)+\mu + \epsilon^{(1)}(t)+\sum_{q=1}^3 \theta^{(q,1)}\epsilon^{(1)}(t-q), \label{eq:kappa1cooked}\\
\kappa^{(2)}(t) &= (1+\phi)\kappa^{(2)}(t-1)-\phi \kappa^{(2)}(t-2) + \epsilon^{(2)}(t)+\sum_
{q=1}^2 \theta^{(q,2)}\epsilon^{(2)}(t-q).
\label{eq:kappa2cooked}
\end{align}

The estimated $ARIMA$ parameters are
$\mu = -0.0195, \phi = 0.9206, \theta^{(1,1)} = -0.5516, \theta^{(2,1)} = 0.1736, \theta^{(3,1)} = 0.5169,
 \theta^{(2,1)} = -1.4664, \theta^{(2,2)} = 0.6167.$  The $\theta^{(q,k)}, k = 1, 2$ describe how past errors echo into future values of $\kappa^{(k)}.$  For example, the large negative value of $\theta^{(2,1)}$ means that the noise generated in $\kappa^{(1)}(s)$ will be amplified, made negative, and added to the future $\kappa^{(1)}(s+2).$  The above equations imply that the state has three components, $Z(T) = \{ \kappa^{(1)}(T), \kappa^{(2)}(T), \kappa^{(2)}(T-1)\}$.

As in the previous case studies, we develop a deterministic estimate for survival probabilities. Denote by $\xi^{(k)}(t,s)  \doteq\E[\kappa^{(k)}(t) \mid Z(s)]$ for $k=1,2$. The expressions for $\xi^{(k)}$ are as follows.
\begin{lemma}\label{lemma:cooked}
The following hold for $t>s$
\begin{align}
\xi^{(1)}(t,s) &= \kappa^{(1)}(s) + \mu(t-s); \\
\label{eq:kappa2cookedexpectation}
\xi^{(2)}(t,s) &= \phi^{t+1-s}\left(\frac{\kappa^{(2)}(s)-\kappa^{(2)}(s-1)}{\phi-1}\right)+ \left(\frac{\phi\kappa^{(2)}(s-1) - \kappa^{(2)}(s)}{\phi-1}\right).
\end{align}
\end{lemma}
The proof can be found in \ref{app:proof3}. Based on Lemma \ref{lemma:cooked}, and substituting expected values of $\kappa^{(k)}(s)$ into \eqref{eq:cbd} we obtain a deterministic estimate of the $u$-year survival probability as the product
%
\[ \hat{P}^{det}(Z(s),t,u,x) = \prod_{j=0}^{u-1} \frac{1}{1+\exp\left(\xi^{(1)}(t+j,s)+ (x+j-{x}_{Ave})\xi^{(2)}(t+j,s)\right)}.\]
Through equation \eqref{eq:life-annuity}, this yields the estimate for the $T-$year deferred annuity given $Z(T):$
\begin{align} \label{eq:cookedannuity}
  \hat{a}^{det}(Z(T);T,x) = \sum_{s=1}^{\bar{x}-x} e^{-rs} \hat{P}(Z(T),T,s,x),
\end{align}
where the cutoff age is $\bar{x}=89.$

We proceed to value life annuities in the above model. In contrast to the first two case studies, we extend the deferral period to twenty years.  An additional ten years of evolution imbues significant uncertainty into the mortality state $Z(T)$. We use an empirical training design $\mathcal{D}$ for this case study for two reasons; one being that the correlation structure is problematic with any organized grid.  From \eqref{eq:kappa2cooked}, we see that $\kappa^{(2)}(20)$ and $\kappa^{(2)}(19)$ should be strongly correlated, while both $\kappa^{(2)}(19)$ and $\kappa^{(2)}(20)$ are independent of $\kappa^{(1)}(20).$  Secondly, the long deferral period causes significant variation in the distribution of $Z(20),$ and with expectation in mind, we desire the accurate capturing of the density of $Z(20)$ that the empirical grid will provide.  The algorithms discussed in Sections \ref{sec:ChoiceOfDesign} and \ref{sec:fitting} are used to generate the design and fit the surrogate models.
As in Section \ref{sec:chen-coxcasestudy}, we choose an ordinary kriging and 1st-order linear universal kriging models, and also fit a thin plate spline model as used in Section \ref{sec:2pop-study}.

\subsection{Results}
\begin{table}[ht]
\begin{center}
\begin{tabular}{rcccc} \hline
& \multicolumn{2}{c}{$N_{tr}=1000$} & \multicolumn{2}{c}{$N_{tr}=8000$} \\
Type & Bias & $\sqrt{\text{IMSE}}$  & Bias & $\sqrt{\text{IMSE}}$ \\ \hline
Analytic & -4.560e-01  &  5.257e-01 & -4.560e-01  &  5.257e-01  \\
Spline & -2.358e-02  &  6.719e-02   & 4.195e-03  &  5.436e-02  \\
Ord.~Kriging & 3.669e-03  &  9.785e-02 &  9.734e-03  &  7.743e-02    \\
Univ.~Kriging & -1.785e-03  &  5.844e-02&  5.635e-03  &  4.355e-02  \\ \hline
\end{tabular}
\end{center}
\caption{Performance of analytic estimates and surrogate models for 20-year deferred annuity values under the CBD framework.  Numbers reported are based on $N_{out} =1000$ draws of $Z(20)$.  $N_{tr}$ is allocated into $N_{tr,1} = N_{tr}^{2/3}$ training points and $N_{tr,2} = N_{tr}^{1/3}$ Monte Carlo batches per training point.
Analytic estimate refers to \eqref{eq:cookedannuity}, and Spline to thin plate spline (TPS) model. Universal kriging model uses linear basis functions. }
\label{tab:CBDresults}
\end{table}

In contrast to the results in Sections \ref{sec:chencoxresults} and \ref{sec:2popresults}, Table \ref{tab:CBDresults} shows that the analytic estimator \eqref{eq:cookedannuity} crumbles under this volatile model and long deferral period.  On the other hand, both kriging models produce reasonable results even with $N_{tr}=1000$.  We can also observe a diminished effect of increasing the training set size, due to the increased model variance.

These results reflect the comments made in the previous sections: the analytic estimate is a parametric guess as to what may provide an accurate result, and that guess is not always correct.  Our analytic choice in this case study was derived along identical lines as to the analytic estimates in the other case studies, yet performs substantially worse.  In comparison, the statistical learning frameworks provide a reliable estimator even in a volatile model with a three-dimensional state process and long deferral period.

\section{Conclusion}\label{sec:conclusion}
The three case studies above showcase the flexibility and admirable performance of the surrogate models across a range of various longevity risk dynamics. Compared to the consistent accuracy of the statistical emulators, the quality of the deterministic projections was widely varying. Because an analytic derivation is required to produce a deterministic estimator, there are several plausible estimators available. In Section \ref{sec:2pop-study} we derived two different estimators, both of which were viable, but one underperformed. Similarly, in Section \ref{sec:cooked} the derived deterministic projection was also inaccurate. Overall, these examples show that our models can outperform deterministic projections and provide minimally-biased estimates.

Relative to the analytic estimates, the case studies in section  \ref{sec:2pop-study} required the training set size to be sufficiently large, while in Sections \ref{sec:chen-coxcasestudy}-\ref{sec:cooked} a small design of less than $10^3$ simulations yielded accurate models. In terms of individual model performance, it was without surprise that the SK model in Section \ref{sec:2pop-study} produced the best results.  The analytic estimate was already performing adequately, and the SK model used it as its trend component to improve accuracy even more. The downside to this is that a well-performing deterministic estimate was required.

Throughout the paper we have hinted at possibilities for further work. Straightforward extensions include using other mortality models, or emulating other insurance products, e.g.~variable life annuities. One can also build more dynamic surrogates that treat initial age $x$ (fixed in our case studies as $x=65$) or deferral period $T$ as part of the state $Z$, providing a joint prediction for $(Z(0),T,x) \mapsto \E[a(Z(T),T,x)]$. Similarly, one could consider more parameter uncertainty which would lead to including additional components in the state $Z$.

The emulators we obtained for $a(Z(T), T, x)$ offer a high-performance tool for annuity risk management. Indeed, they are based on advanced, previously vetted stochastic mortality models and calibrated to real, reliable, large-scale mortality datasets. Hence, the fitted estimates for annuity values are in essence a best-available forecast that combines state-of-the-art longevity modeling, data calibration and statistical model. As such, (after incorporating age and interest rate as model parameters) they would be of independent interest to actuaries working in longevity space and seeking easy-to-use tools for forecasting net present values of life annuities. The emulator offers a plug-and-play functionality, converting inputted parameters (such as age $x$, deferral period $T$ and discount rate $r$) into the annuity value (note that the initial state $Z(0)$ is read off from the calibration procedure).  One can imagine building a library of such emulators for different mortality-contingent products available in the marketplace.

Looking more broadly, the emulation approach we propose is very general and can be applied in a variety of actuarial contexts. In particular, in future work we plan to extend it to the microscopic agent-based models of mortality \citet{Barrieu2012} which offer a canonical ``complex system'' representation of population longevity. We believe that emulators could significantly simplify predictions in these types of models by providing a tractable, statistical representation of demographic interactions within a stochastic dynamic population framework. Another class of insurance applications requires functional-regression tools where emulators can again be very effective \cite{LinGan15}. A different extension is emulation of risk measures related to $F(T,Z(\cdot))$, such as VaR or TVaR, which require targeted surrogates that focus on a specific region of the input space. A starting point is to combine concept of importance sampling to generate a targeted design $\mathcal{D}$ that e.g.~preferentially concentrates on the left tail of $F$.


\bibliographystyle{elsarticle-harv}
\bibliography{mybib}

\newpage
\appendix
\section{Lee Carter \& CBD Stochastic Mortality Models} \label{sec:leecarter-cbd}
In this section we give a brief summary of existing stochastic mortality models. We use the notation of
\citet{cairns2011mortality} who provided a comprehensive comparison of several mortality models using CMI data.

The APC Lee-Carter model (introduced by \citet{renshaw2006cohort}) models the log mortality rate as
\[ \log m(t,x) = \beta^{(1)}(x) + \beta^{(2)}(x) \kappa^{(2)}(t) + \beta^{(3)}(x) \gamma^{(3)}(t-x). \tag{M2}\]
One can interpret $\beta^{(1)}(x)$, $\kappa^{(2)}(t)$ and $\gamma^{(3)}$ as the age, period and cohort effects, respectively.  The original model proposed by \citet{lee1992modeling} is a special case where $\gamma^{(3)}=0.$  The age effects $\beta^{(k)}(x), k=1,2,3$ are estimated (non-parameterically) from historical data, while the period and cohort effects are taken as stochastic processes. In the original proposal in \cite{lee1992modeling}, the period effect $\kappa^{(2)}$ is assumed to follow a random walk (i.e.~unit root $AR(1)$ in discrete time),
\[ \kappa^{(2)}(t) = \kappa^{(2)}(t-1) + \mu^{(2)} + \sigma^{(2)} \eps^{(2)},\]
where $\mu^{(2)}$ is the drift, $\sigma^{(2)}$ is the volatility, and $\eps^{(2)} \sim N(0,1)$ i.i.d.~is the noise term. Alternatively, \citet{cairns2011mortality} mention that $ARIMA$ models may provide a better fit, in particular fitting an $ARIMA(1,1,0)$ process for $\kappa^{(2)}$ based on 2007 CMI dataset.

For the cohort effect, \citet{renshaw2006cohort} suggested using $ARIMA$ models for $\gamma^{(3)}(t-x)$; \citet{cairns2011mortality} recommend the use of either $ARIMA(0,2,1)$ or $ARIMA(1,1,0)$.  \citet{renshaw2006cohort} and \citet{cairns2011mortality} both assume $\gamma^{(3)}$ is independent of $\kappa^{(2)}.$

This model has identifiability issues, and one set of constraints could be
\[ \sum_t \kappa^{(2)}(t) = 0, \quad \sum_x \beta^{(2)}(x) = 0, \quad \sum_{x, t} \gamma^{(3)}(t-x) = 0, \quad \text{ and } \quad \sum_x \beta^{(3)}(x) = 1.\]


From a different perspective, \citet*{cairns2006two} (CBD) proposed a model for $q(t,x)=1-P(Z(0); t, 1, x)$, the probability of death in year $t$ for someone aged $x$. Namely, they use
\[ \logit q(t,x) = \beta^{(1)}(x)\kappa^{(1)}(t) + \beta^{(2)}(x) \kappa^{(2)}(t), \tag{M5}\]
where $\logit(y) = \log \left(\frac{y}{1-y}\right)$. 

If we let $n_a$ be the number of ages available in the data set for fitting, and take ${x}_{Ave} = n_a^{-1} \sum_i x_i$, the commonly used parameterization for the  CBD model (M5) is
\begin{equation}
\beta^{(1)}(x)=1, \qquad \text{ and } \quad \beta^{(2)}(x) = x-{x}_{Ave}.
\end{equation}
Under these assumptions there are no identifiability issues.

\section{Proofs of Analytic Estimates}\label{sec:proofs}
\subsection{Proof of Lemma \ref{lemma:chen-cox}.}
Since the noise terms $\xi^{(k)}(u)$ are independent of $\kappa(s)$ for $u \neq s$, taking conditional expectation with respect to $Z(s)=\{\kappa^{(1)}(s), \xi^{(2)}(s)\},$ and writing in terms of the increments $\kappa(u)-\kappa(u-1)$  yields
\begin{align} \notag
\E\left[\kappa(t)-\kappa(s)\mid Z(s)\right] &=  \sum_{u=s+1}^t \E\left[ \kappa(u)-\kappa(u-1)\mid Z(s)\right]\\
	&=  \sum_{u=s+1}^t \E\left[\xi^{(1)}(u) + \xi^{(2)}(u) - \xi^{(2)}(u-1) \mid Z(s)\right]. \label{eq:proof1eq2}
\end{align}
By the independence assumption we have for $u \neq s+1$ 
\begin{align}
\E\left[\xi^{(1)}(u) \mid Z(s)\right] & = \mu^{(1)} \\
\E\left[\xi^{(2)}(u) - \xi^{(2)}(u-1)\mid Z(s)\right]  & = \mu^{(2)}p - \mu^{(2)}p =0.
\label{eq:proof1eq3}
\end{align}
For $u = s+1,$
\begin{align}
\E\left[\xi^{(2)}(s+1) - \xi^{(2)}(s)\mid Z(s)\right] = \mu^{(2)}p - \xi^{(2)}(s). \label{eq:proof1eq4}
\end{align}
Combining \eqref{eq:proof1eq2}-\eqref{eq:proof1eq4}, we obtain
\begin{align}
\E\left[ \kappa(t) \mid Z(s)\right] &= \kappa(s) + (t-s)\mu^{(1)} + \mu^{(2)}p - \xi^{(2)}(s).
\end{align}
\qed

\subsection{Proof of Lemma \ref{lemma:two-pop}.}

Since $\kappa_1$ has trend $\mu_1$, $\E[\kappa_1(t)-\kappa_1(t-1)] = \mu_1$, and using conditional independence, we obtain,
\begin{align}
\E\left[\kappa_1(T+t) \mid Z(T)\right] 
	&= \kappa_1(T) + \mu_1 t. \label{eq:proof2-E(kappa1)}
\end{align}
For the co-integration term $S(t)$, the expected values satisfy
\begin{equation}
\E[S(T+t) \mid Z(T)] = \mu_2 + \phi\left(\E[S(T+t-1)\mid S(T)]-\mu_2\right). \label{eq:proof2-recursive}
\end{equation}
The above gives a recursive equation for $t \mapsto \E[S(T+t) \mid Z(T)]$, with initial condition $\E[S(T+0)\mid Z(T)]=S(T),$ which can be solved to yield
\begin{equation}
\E[S(T+t)\mid Z(T)] = \mu_2(1-\phi^t)+\phi^tS(T). \label{eq:proof2-ES(T)}
\end{equation}
Finally, using $\kappa_2(t) = \kappa_1(t)-S(t),$ and combining \eqref{eq:proof2-E(kappa1)} with \eqref{eq:proof2-ES(T)} leads to
\begin{align*}
\E[\kappa_2(T+t)\mid Z(T)] 
	&=  \kappa_1(T) + \mu_1 t -\left(\mu_2(1-\phi^t)+\phi^t \left[\kappa_1(T)-\kappa_2(T)\right]\right). \\
\end{align*}
as desired. \qed\\

\subsection{Proof of Lemma \ref{lemma:cooked}}\label{app:proof3}

For $\xi^{(1)}(t,s)$, $\kappa^{(1)}$ is no different than a random walk with drift, so we have
\[\E[\kappa^{(1)}(t)\mid \kappa^{(1)}(s)] = \kappa^{(1)}(s) + \mu(t-s), \quad s \leq t.\]
Next, we take expectation on both sides of \eqref{eq:kappa2cooked} to obtain the recursive relation
\begin{equation}\label{eq:proof3recursive}
\E[\kappa^{(2)}(t) \mid Z(s)] = (1+\phi)\E[\kappa^{(2)}(t-1) \mid Z(s)] - \phi \E[\kappa^{(2)}(t-2) \mid Z(s)]
\end{equation}
where $Z(s) = \{\kappa^{(1)}(s), \kappa^{(2)}(s), \kappa^{(2)}(s-1)\}$.  Equation \eqref{eq:proof3recursive} is a recursive relation in $t$ with general solution
\begin{equation}\label{proof3solution}
\E[\kappa^{(2)}(t) \mid Z(s)] = c_1\phi^t+c_2,
\end{equation}
where the constants $c_1$ and $c_2$ are to be determined.  Plugging-in the initial conditions
\begin{align}
c_1\phi^s + c_2 &= \E[\kappa^{(2)}(s) \mid Z(s)] = \kappa^{(2)}(s),\label{eq:proof3.1} \quad\text{and}\\
c_1 \phi^{s+1}+c_2 & = \E[\kappa^{(2)}(s+1) \mid Z(s)] = (1+\phi)\E[\kappa^{(2)}(s)  - \phi \kappa^{(2)}(s-1) \mid Z(s)] \notag\\
	&= (1+\phi)\kappa^{(2)}(s) - \phi\kappa^{(2)}(s-1).\label{eq:proof3.2}
\end{align}
and solving for $c_1, c_2$ we obtain
\begin{equation}\label{eq:proof3c1c2}
c_1 = \phi^{1-s}\frac{\kappa^{(2)}(s)-\kappa^{(2)}(s-1)}{\phi-1}, \qquad c_2 = \frac{\phi\kappa^{(2)}(s-1) - \kappa^{(2)}(s)}{\phi-1}.
\end{equation}
Finally, combining \eqref{eq:proof3c1c2} with \eqref{proof3solution}, we arrive at \eqref{eq:kappa2cookedexpectation}.
\qed

\end{document}